# Claws, Disorder, and Conformational Dynamics of the C-terminal Region of Human Desmoplakin

Charles E. McAnany and Cameron Mura[*]

*Department of Chemistry, University of Virginia, Charlottesville, VA 22904, USA*

E-mail: cmura@muralab.org

Phone: 1.434.924.7824. Fax: 1.434.924.3710

---

[*]To whom correspondence should be addressed

**Abstract**

Multicellular organisms consist of cells that interact via elaborate adhesion complexes. Desmosomes are membrane-associated adhesion complexes that mechanically tether the cytoskeletal intermediate filaments (IFs) between two adjacent cells, creating a network of tough connections in tissues such as skin and heart. Desmoplakin (DP) is the key desmosomal protein that binds IFs, and the DP•IF association poses a quandary: desmoplakin must stably and tightly bind IFs to maintain the structural integrity of the desmosome. Yet, newly synthesized DP must traffick along the cytoskeleton to the site of nascent desmosome assembly without 'sticking' to the IF network, implying weak or transient DP⋯IF contacts. Recent work reveals that these contacts are modulated by post-translational modifications (PTMs) in DP's C-terminal tail. Using molecular dynamics simulations, we have elucidated the structural basis of these PTM-induced effects. Our simulations, nearing 2 μs in aggregate, indicate that phosphorylation of S2849 induces an 'arginine claw' in desmoplakin's C-terminal tail ($DP_{CTT}$). If a key arginine, R2834, is methylated, the $DP_{CTT}$ preferentially samples conformations that are geometrically well-suited as substrates for processive phosphorylation by the cognate kinase GSK3. We suggest that $DP_{CTT}$ is a molecular switch that modulates, via its conformational dynamics, DP's efficacy as a substrate for GSK3. Finally, we show that the fluctuating $DP_{CTT}$ can contact other parts of DP, suggesting a competitive binding mechanism for the modulation of DP⋯IF interactions.

# Introduction

**Desmosomes mediate cellular adhesion**— Desmosomes are inter-cellular junctions found in epithelial and cardiac tissue.[1–7] By connecting the intermediate filaments (IFs) of neighboring cells, desmosomes create a network of adhesive structural interactions that impart tensile strength and durability to these tissues. The general architecture of the desmosome is shown in Figure 1. Desmosomes expose the extracellular regions of two transmembrane cadherins, desmocollin and desmoglein, on the cell surface; these proteins bind the cadherins of neighboring cells via $Ca^{2+}$-dependent homo- or heterophilic interactions. The desmosomal cadherins traverse the plasma membrane and bind two other key proteins, plakoglobin and plakophilin, which in turn bind a large, essential protein known as desmoplakin (DP; Figure 1). DP binds to the cytoskeletal IFs and, because the cytoskeleton spans the cytosol of one cell and binds to other desmosomes (which in turn bind to other, neighboring cells), this extended network of adhesive molecular contacts links together cells into tissues.

The IFs bind to DP's three plakin repeat domains (PRDs), which correspond to residues 1960-2208 and are denoted PRD A, PRD B, and PRD C (Figure 1).[2,8] The C-terminal PRDs are connected to the plakin domain, in DP's N-terminal region, via a fibrous rod (residues 1057-1945, central coiled-coil in Figure 1). The coiled-coil region is responsible for DP dimerization and, ultimately, links an electron-dense region known as the *outer dense plaque* (near the cell membrane) to the *inner dense plaque* (proximal to the IF network), across a span of $\approx$ 10-20 nm (Figure 1).[4,9] The plakin domain of spectrin repeats (residues 178-883, leftmost structure in DP in Figure 1)[10] provides a relatively rigid N′-terminal connection that binds to the plakophilin (PKP) and plakoglobin (PG) proteins, thereby helping target DP to the desmosome.[11] Plakoglobin binds to the intracellular regions of desmocollin and desmoglein, denoted as the cadherin cytoplasmic regions (CCR) in Figure 1. Crystallographic structures of PRDs have revealed a basic groove that can sterically accommodate IFs, suggesting that as a potential mode of DP⋯IF interactions.[8,10,12,13]

Because desmosomes impart structural integrity and mechanical strength to cell⋯cell junctions, aberrant desmosome function underlies several diseases of the skin and heart.[7] For example, pemphigus is an autoimmune disease caused by antibodies to desmoglein,[14] the DP mutation S2594P is linked to Carvajal syndrome,[15] and several DP mutations are associated with the lethal heart disease arrhythmogenic right ventricular cardiomyopathy.[15–17] Downregulation of DP has been linked to metastasis of tumor cells,[18] and desmosome function in cancer remains an active area of research.[2] Several point mutations in the desmoplakin C-terminal tail ($DP_{CTT}$) have been examined previously,[19–21] providing evidence that the $DP_{CTT}$ region regulates DP•IF adhesion.

Include Figure 1 here.

Cellular adhesion by desmosomes is regulated by two principal mechanisms: (i) $Ca^{2+}$-dependent adhesion of extracellular cadherin domains[21] and (ii) phosphorylation-dependent adhesion of DP to IFs.[7] During desmosome formation, DP must be translocated to the desmosome along the cytoskeletal network, and therefore must bind only loosely to IFs. Once DP reaches the desmosome and is properly localized, it binds more tightly to the IFs in order to create stable and persistent intercellular connections in epithelial tissues (e.g., skin) and cardiac muscle. The IF-binding site of DP is required for normal desmosome assembly *in vivo*, suggesting that DP transport occurs along IFs;[22] however, the S2849G mutation, which is in the $DP_{CTT}$, causes DP to associate abnormally strongly with IFs, thereby retarding desmosome assembly.[22]

**Post-translational modifications alter the behavior of desmoplakin**— Cell biological and proteomic work suggest that assembly of the DP•IF adhesion complex is regulated by specific post-translational modifications (PTMs) in DP, including phoshporylation at S2849 in the $DP_{CTT}$.[7,19] At least two kinases are suspected to phosphorylate DP: protein kinase C-$\alpha$ (PKC$\alpha$) and glycogen synthase kinase 3 (GSK3). PKC$\alpha$ binds to DP in the cytoplasm and phosphorylates $DP_{CTT}$ to initiate desmosome assembly.[23] Once S2849 is phosphorylated, a second kinase, GSK3, further phosphorylates $DP_{CTT}$ in a processive manner.[20] GSK3 is a processive kinase that recognizes peptides with the sequence $SXXXS_{PO_3}$ and phosphory-lates the serine.[24] Suitable substrates for GSK3 are generally those peptides that have been already phosphorylated, and the processive phosphorylation cascade proceeds in a $C'\to N'$ direction.[24] While its cellular activity is regulated by various factors, GSK3 does not display strong substrate specificity on its own.[25,26] Recent *in vivo* studies[20,27] have shown that the S2849G mutation has the same effect on DP as does inhibiting PKC$\alpha$ or GSK3—namely, DP binds IFs tightly as soon as DP is synthesized, slowing its recruitment to the assembling desmosome. The site of these phosphorylation events (i.e., the DP$_\text{CTT}$) features a glycine/serine/arginine-rich region (GSRR) containing the sequence (GSRS)$_5$GSRRGS.

In addition to the S2849G mutation, which exhibits deleteriously enhanced IF binding, an R2834H mutation causes cardiac dysfunction in mice, and various other mutations in DP$_\text{CTT}$ are linked to various disease states.[27] The residue R2834 in the DP$_\text{CTT}$ is important because its dimethylation (giving R$_{\text{Me}_2}$2834) may serve as a molecular switch for the processive phosphorylation of this region. Indeed, DP$_\text{CTT}$ is phosphorylated at multiple sites, and this phosphorylation cascade is contingent on two PTMs: a phosphorylation (S$_{\text{PO}_3}$2849) and a methylation (R$_{\text{Me}_2}$2834). Because the R2834H mutation clearly precludes the R$_{\text{Me}_2}$2834 state, this mutant DP$_\text{CTT}$ is phosphorylated only at S$_{\text{PO}_3}$2849, and therefore presumably binds tightly (and gets stuck) to the IF network.[20,27]

As mentioned above, GSK3 generally exhibits low substrate specificity. The R2834H mutation provides an interesting counterpoint to this trend. Since GSK3 binds DP$_\text{CTT}$ at S$_{\text{PO}_3}$2849 in order to phosphorylate S2845, it is surprising that a relatively minor change (a point mutation), eleven residues away from S2845, prevents GSK3 from initiating processive phosphorylation.[27] Therefore, the DP$_\text{CTT}$ also provides a useful system to explore the structural and dynamical basis of GSK3 substrate recognition.

In addition to phosphorylation, mass spectrometry (MS) studies of DP$_\text{CTT}$ have revealed multiple methylated arginine residues, with up to six methyls concurrently in the DP$_\text{CTT}$.[20,27] (In that case, three of the seven arginine residues in the DP$_\text{CTT}$ were dimethylated.) In cases where a single arginine is dimethylated, some evidence indicates that both methyls are on the same nitrogen, yielding an asymmetric dimethylarginine residue.[28] In $DP_{CTT}$, methylation appears to be necessary before GSK3 can initiate processive phosphorylation. Since their initial discovery in the 1960s,[29] methylated protein residues often have been found to occur in serine-rich region (SRR) regions; indeed, such sequences serve as a common substrate for methyltransferases.[30–32] However, unlike phosphorylation, methylation is not known to be metabolically reversible, at least not outside the context of histones.[33] Apart from regulating the phosphorylation cascade of $DP_{CTT}$, any functional roles of these methylations remain unexplored.

**Arginine claws can structurally rigidify disordered regions**— The $DP_{CTT}$ contains an SRR which is multiply phosphorylated,[27] but any structural and dynamical effects of PTMs in the $DP_{CTT}$ remain unknown. The structural dynamics of heavily-phosphorylated SRRs have been studied in other systems, and phosphorylation of SRRs is a common regulatory mechanism in the Eukarya.[34] A three-dimensional (3D) structure known as the *arginine claw* provides a rationale for some of these interactions and effects.

The arginine claw (RC), a relatively recently-identified structural element of SRRs, was first characterized[35] in the C-terminal region of ASF/SF2, a protein involved in mRNA splicing, spliceosome assembly, and mRNA nuclear trafficking.[36] This protein is phosphorylated in an SRR, and this modification serves as a nuclear import signal. Fundamentally, the compaction of a peptide region into an RC sequesters charged side-chains away from the protein surface (Figure 2). Implicit-solvent molecular dynamics (MD) simulations of a fully-phosphorylated $(RS_{PO_3})_8$ peptide[35] initially revealed a compact structure, with one phosphate group coordinated by the guanidinium moieties of several arginine residues. An RC such as we find in the $DP_{CTT}$ (see below) is shown in Figure 2c, alongside an illustration of the RC originally characterized by Hamelberg et al.[35] in Figure 2d. Such structures as shown in Figure 2d were found to stably persist over the 200-ns, fully-atomistic, explicit-solvent MD simulations of the $(RS)_8$ system.[35] In multiply-phosphorylated SRRs, those phosphate groups not involved in the RC are solvent-exposed, and this dynamically-varying surface exposure has been proposed as the recognition mechanism for nuclear import of a serine/arginine–rich ASF/SF2 (this particular 'SR protein' is also known as SRSF1).[35] NMR studies of the ASF/SF2 system, as well as hPrp28 (another RNA-splicing–related system), have complemented the results of MD simulations, demonstrating that the phosphorylation of SRRs rigidifies the region.[37] Further simulation-based studies of RCs showed that claw formation allows the SRR of the lamin B receptor to bind to histones, despite the large positive charges of both interacting proteins.[38] Crystallographic studies of the RNA splicing factor SF1 have also revealed a partial RC.[39] As a final recent example, simulations have detected a claw-like structure in the long-time dynamics of a small, apoptosis-related intrinsically disordered protein (IDP) known as Noxa.[40]

Include Figure 2 here.

**Simulations of disordered structural ensembles are not straightforward**— MD simulations[41,42] have been used to examine SRRs, IDPs, PTMs and, to a lesser extent, the interplay between these.[37,43–51] The long timescales of conformational transitions and structure formation in SRRs has often prompted the use of relatively inexpensive implicit-solvent models. However, continuum solvent models likely overestimate the electrostatic effects of salt bridges in determining three-dimensional structure,[52] and RC simulations performed with implicit solvent models predict more compact structures than do analogous explicit solvent simulations.[35] Another important consideration is the force-field (FF) used to describe the potential energy landscape of a system. Modern FFs have been used to predict protein structures, albeit with limited success;[53] any FF shortcomings are exacerbated in simulations of IDPs due to the small energy differences between conformations.[54] Recent work has shown that CHARMM36 and ff03* predict substantially different secondary structures in glycosylated IDPs.[50] Simulations of highly-charged systems are also affected by the inadequate representation of electronic polarizability in current FFs. The classical Coulomb model of electrostatic interactions has been extended to include polarizability, though polarizable FF parameters are not yet available for PTMs such as in the systems studied here.[55]

FFs are generally parameterized against the physicochemical properties of well-characterized model systems, for which experimental data or high-level quantum mechanical calculations are available.[56,57] Disordered peptides are often underrepresented in these parameterization processes, as validating a structural ensemble generated by simulations of an IDP may be experimentally challenging (versus non-IDP systems).[51] Not only are structural parameters difficult to determine experimentally,[58] trajectory analysis is seldom straightforward and many complex techniques have been employed in analyzing IDP simulation results.[59] Finally, note that the RC is a somewhat unusual system insofar as it has a highly-charged core, while FFs are parameterized against the more common cases wherein charged residues are solvent-exposed. For these reasons, we note that simulations of systems of this type should be considered more suggestive and predictive rather than conclusive.

For slow processes and rare events, the computational cost of simulating a system such as an IDP for a sufficient length of time may be untenable. Several enhanced sampling methods have been developed.[60–62] However, the size of the $DP_{CTT}$, with its extended starting conformation (and requisite number of solvent molecules; Figure 2a), necessitates a large number of replicas for replica-exchange simulations, and correspondingly long trajectories are required for adequate mixing[63] of the replicas (McAnany & Mura, data not shown).

**Our MD simulations of DP**— We used classical, all-atom MD simulations to examine the structural effects of PTMs in the $DP_{CTT}$, with a specific aim of elucidating the conformational dynamics of this 70-residue region (Figure 1) and the riddle of strong/weak DP⋯IF interactions (might $DP_{CTT}$ be a PTM-modulated molecular switch?). To mitigate the effects of FF inaccuracies and limited sampling, each system was simulated under two independent FFs (from the Amber and CHARMM families), and each production trajectory is at least 100-ns long. Simulations were extended to 200 ns for all phosphorylated systems; for consistency in scaling the figures, the 200-ns simulations were split into 100-ns chunks. When we refer to a simulation without explicitly mentioning a time, we refer exclusively to the first 100 ns; when referring to the second 100 ns, we call this 'cycle2'.

We begin by proposing a quantitative definition of an RC, and we show that simultaneous methylation and phosphorylation cause $DP_{CTT}$ to assume conformations that are compatible with GSK3–binding. We propose that $DP_{CTT}$, in its various claw and non-claw states, competes with IF molecules for binding sites on the neighboring PRD elements (Figure 1), thus suggesting a straightforward dynamical mechanism for the regulation of DP⋯IF interactions.

# Methods of Procedure

## Molecular dynamics simulations

Classical, all-atom MD simulations were performed using NAMD 2.9,[64] with either the Amber PARM99SB[65] or CHARMM36[66,67] force-field. Parameters for modified residues, such as diprotic phosphoserine ($S_{PO_3}$), were drawn from[68] and[69] as available (see below). No crystallographic or NMR structure of the $DP_{CTT}$ is available, so the peptide $^{2802}$`LLEAASVSS KGLPSPYNMS SAPGSRSGSR SGSRSGSRSG SRSGSRRGSF DATGNSSYSY SYSFSSSSIG H`$^{2871}$ was constructed using VMD's `Molefacture` plugin in protein builder mode (VMD v1.9.1);[70] note that the above sequence numbering matches human DP (UniProt ID P15924), and the simulated $DP_{CTT}$ peptide ends at the very $C'$-terminus of DP. The peptide was constructed in an extended conformation ($\phi = 180°, \psi = 180°$), as shown in Figure 2a. PTMs were applied to specific residues (Figure 1) by using either leap (for PARM99SB, LEaP from `AmberTools13`[65]) or patches in VMD's `psfgen` tool (for CHARMM36). Each initially-extended peptide system was subjected to a brief conformational relaxation simulation in implicit solvent. These relaxation simulations were performed with rigid hydrogen atoms, a nonbonded cutoff distance of at least 11.0 Å, and a Langevin thermostat set to human physiological temperature (310 K). NAMD's generalized Born implicit solvent model[71] was used with an ion concentration of 0.15 M. A 2-fs integration timestep was used in all simulations. The relaxation simulation consisted of 10,000 steps of conjugate gradient potential energy minimization, followed by 10 ns of unrestrained MD. A representative relaxed structure is shown in Figure 2b.

Periodic boundary conditions were set-up by solvating the final structures from the relaxation simulations in a truncated octahedral cell of water molecules, of sufficient dimensions such that there would be at least 15 Å of water between the peptide and the envelope of the cell (this worst case scenario being reached if the peptide were to adopt the most extended state found in the last 5 ns of the relaxation simulation). This heuristic was adopted because of the periodic boundary conditions used in the explicit-solvent simulations: the peptide will be flexible during the production runs, and any prolonged violation of a 30 Å distance between periodic images of the $DP_{CTT}$ solute could introduce artifacts. To mitigate computational costs, a "worst case" expanded size for the peptide was estimated based on the last half of the relaxation run; the first half of the relaxation run was not used in our geometry calculations, as the peptide is still collapsing during that time from its initial (extended) state. Even with the relaxation simulation, most of our simulated systems still contained over 200,000 particles (mostly $H_2Os$). Waters were placed about the compactified peptide using the SOLVATE program,[72] with custom modifications introduced in-house to enhance its performance. Ions were placed by VMD's `Autoionize` plugin (for CHARMM36) or LEaP (for PARM99SB) to reach 0.15 M NaCl. Because LEaP's ion placement was observed to be non-random, a 10-ns water equilibration run was performed on those systems simulated using PARM99SB; this run comprised 100 steps of energy minimization, followed by 10 ns of dynamics with the protein atoms harmonically restrained by a force constant of 1 kcal/mol/Å$^2$. All other parameters were the same as in the equilibration runs.

For consistency, all PARM99SB and CHARMM36 systems were equilibrated in the same way, using the general approach of Mura & McCammon.[73] Again, a 2-fs timestep, with at least an 11.0 Å nonbonded cutoff and a 310 K Langevin thermostat, were used. Periodic boundary conditions were employed with particle mesh Ewald (PME) electrostatics and a grid spacing of better than 1/Å per direction. NAMD's langevinPiston feature was used to maintain pressure at 1 atm. Protein atoms were initially harmonically restrained by a 50 kcal/mol/Å$^2$ spring. The systems were minimized for 1000 steps, then gradually heated in 10 K increments, with 2 ps of dynamics at each new temperature. Once the system temperature reached 310 K, the restraints were weakened to 0.01 kcal/mol/Å$^2$ by repeatedly halving the restraint strength and simulating for 2 ps. Finally, the restraints were completely removed and the system was equilibrated for 10 ns in the NPT ensemble.

Production trajectories were computed using the same simulation parameters as the equilibration runs described above, and were extended to at least 100 ns each (Table 1). Analyses were performed using VMD and custom scripts written in the Python[74] and D[75] languages. All simulation and analysis scripts are available upon request, as are dehydrated trajectories.

## Methylarginine parameterization

Parameters for dimethylarginine, $R_{Me_2}$, in the CHARMM family of FFs were generously contributed by the Dejaegere laboratory.[76] These parameters lacked a term for the CK1–NH1–CK2 angle, subtended by the carbons of the added methyl groups and the nitrogen to which they are bonded; therefore, the value of this term was estimated using *ab initio* quantum mechanical calculations on a single $R_{Me_2}$ residue. Specifically, the GAMESS[77,78] program was used to perform geometry optimizations at the RHF/3-21G level in implicit water.[79] First, the optimal equilibrium geometry was determined, then the relevant bond angle was constrained 1° higher than the equilibrium angle and the equilibrium geometry re-calculated subject to this constraint. The derivative of energy with respect to angle provides the necessary value for this new FF parameter. The angle constraint was found to be 95.467 kcal/mol/rad$^2$, with an equilibrium angle value of 115.252°.

## Analysis pipeline

For the sake of data-processing consistency, comparability, and automation, software tools were developed into a pipeline to analyze each simulation trajectory in a standardized manner. The detailed results of these analyses are shown in Figures S1 to S19. Figures were prepared using matplotlib and Python 3.3, with some analysis steps performed in VMD and the D programming language. Detailed descriptions of our analysis modules follow in the remaining subsections.

**Arginine clawicity, $Cy_*^R$ (panel a)**— Plots of the arginine clawicity, arginine clawicity ($Cy_*^R$), show, at each trajectory time-step, the $Cy_*^R$ of the simulated system. For each residue in the sequence, the number of hydrogen bonds made to arginine are calculated, and the largest of these numbers (the $Cy_*^R$, by definition, where the '*' wildcard denotes any residue) is plotted as a blue point. A green trace, representing a 1-ns running average, smoothens the noisy behavior of $Cy_*^R$. On the right of the panel, a vertically-oriented histogram shows the distribution of $Cy_*^R$ values over the entire simulation; an example is given in Figure 3a. These histograms (e.g., ) are also used within the text to succinctly convey the $Cy_*^R$ behavior of a given simulation.

**Residue-specific arginine clawicity, $Cy^R$ (panel b)**— Plots of residue-specific arginine clawicity ($Cy^R$) show which residues are contained in an RC, as exemplified in Figure 3b. For each residue, at each time-step, the number of hydrogen bonds to arginine is calculated. These data are averaged with a 1-ns window before plotting, in order to avoid aliasing. White areas indicate that no hydrogen bonds were made to arginine by a particular residue at a particular time. For clarity, the $DP_{CTT}$ sequence is staggered (up/down) along the horizontal axes of these plots: Residues on the top line align with inward-facing ticks and residues on the bottom line align with the extended outward-facing ticks. The key residues H2834 and S2849 are marked with asterisks.

**SASA of residues 2849 and 2834 (panels c and d)**— Solvent-accessible surface areas were calculated using VMD's SASA tool, with a solvent probe radius of 1.4 Å. As with $Cy^R_*$, the solvent-accessible surface area (SASA) for each frame is shown as a blue point and a green trace shows a 1-ns running average. SASA values were calculated for the entirety of a residue, so comparison between systems with different residue modifications or mutations requires caution, as the residues are of different size. The histogram adjoined to the right axis (200 bins) shows the distribution of SASA over the entire simulation.

**The S2849·····S2845 distance (panel e)**— For each frame in the simulation, the distance between the hydroxyl oxygens of S2849 and S2845 was calculated and plotted as a blue point. The green trace shows a 1-ns running average, and the histogram on the right (200 bins) shows the distribution of distances for the entire simulation.

**GSK3 clash scores (panel f)**— The GSK3 steric clash scores were evaluated via what effectively became a one-dimensional docking procedure (Figure 5). We began with the 3D structure of GSK3, taken as chain A from PDB entry 1I09.[24] The (side-chain) oxygen of residue 338 of chain B (the *recognition site* landmark), and the solvent-facing oxygen of the phosphate docked to chain A (the *active site* landmark), were used as reference points for alignment. These two reference points correspond to the recognition site and active site of GSK3. Note that only those chain A protein atoms built into the crystal structure were considered in the evaluation of clash scores. The corresponding pair of atoms from DP are the side-chain oxygens of S2849 (phosphorylated prior to GSK3 interaction) and S2845 (destined for phosphorylation by GSK3). In phosphorylated systems, the oxygen attached to the carbon was used. For each frame of each trajectory, DP and GSK3 were aligned based on the two pairs of atoms described above. GSK3 was then rotated, in 1° increments, about the axis defined from these four reference points. For each configuration, the number of clashes was taken as the number of contacts between atoms in GSK3 and atoms in DP (sans hydrogens for computational efficiency), with a 2 Å sweep radius. The minimum number of contacts, considered among all rotated positions for the trajectory frame in question, is defined as the *clash score* for that frame; it is this quantity which is plotted in the panels f.

**Contact maps (panel g)**— The contact map shows the pairwise contacts within a protein 3D structure, measured as a symmetric matrix of interatomic distances, $d_{i,j}$, for all pairs of residues $i$ and $j$. The distance is defined so as to account for side-chain interactions: for a given residue pair, all pairs of atoms within each of the two residues $(i_x, j_y)$ are considered, where atom $x$ $(i_x)$ is from residue $i$ and atom $y$ $(j_y)$ is from residue $j$. The contact map distance for $(i, j)$ is then taken as the distance between the closest pair of atoms for all of those pairs within the residue pair. In our illustrations, the lower-left triangle of the contact map shows the *average inter-residue distance* for the duration of the simulation, while the upper triangle gives the *minimum distance* considered over the entire trajectory. The horizontal axis is identical to that used for $Cy^R$, and the vertical axis is marked every ten residues and at the residues that were PTM sites in this study (asterisks).

**Ramachandran plots (panel h)**— Ramachandran plots show the distribution of peptide backbone torsion angles, $(\phi, \psi)$, for each system, along the entire trajectory. Colors are graded by the logarithm of the probability density of a given $(\phi, \psi)$ configuration. Regions corresponding to canonical secondary structures are demarcated by guidelines, with the boundaries drawn from the MolProbity source code.[80] The percent of observations in each region is given at the top of the panel, and these regions roughly correspond to secondary structures: 'L$\alpha$' = left-handed $\alpha$-helix; 'L$\alpha$+' = generously-allowed left-handed $\alpha$-helix; 'e' = $\epsilon$-turn regions, often found ahead of a helix or strand; '$\alpha$' = standard (right-handed) $\alpha$-helix; '$\beta$' = $\beta$-strand; 'g+' = generously-allowed helix or strand; 'o' = other structures.

# Results

## Arginine claws occur in the DP$_{\text{CTT}}$

**A claw can be quantitatively defined, and occurs in the DP$_{\text{CTT}}$**— Past efforts have qualitatively detected RCs based on visual analysis of trajectories, such as the one shown in Figure 2d.[35,38] These past claws (i) were characterized as multiple arginines interacting with a phosphate, (ii) were found to be stable on the timescale of a 100-ns simulation, and (iii) had estimated free energies of formation of $\approx$ –5 kcal/mol.[35] While those attributes describe the behavior of a claw, they are not suitable metrics for determining the claw-forming propensity across a number of trajectories, which is a goal in our current study. First, the above set of descriptors does not, in and of itself, provide an algorithmic solution to the decision problem of whether a particular structure is or is not an RC. Second, the above description involves kinetic and thermodynamic information, both of which require more computationally expensive calculations than would a straightforward geometric definition of an RC. Finally, the above description of a claw does not work well for a trajectory that transiently adopts a claw or claw-like conformation. Therefore, we propose a definition of a claw that is akin to that of a protein secondary structural element.

Our definition is purely geometric, based only on a definition of the hydrogen bond,[81,82] and our parameter is easily evaluated for an arbitrary 3D structure. We define the *clawicity*, $\text{Cy}_{\text{B}}^{\text{A}}$, as the maximum number of hydrogen bonds made by any residue in B to all residues in A. For example, $\text{Cy}_{\text{S51}}^{\text{R}}$ refers to the number of hydrogen bonds made by S51 to all arginine (R) residues. $\text{Cy}_{\text{S}}^{\text{R}}$ refers to the number of hydrogen bonds made to an arginine by the serine (*any* serine) with the greatest number of hydrogen bonds to arginine. We define the *arginine clawicity* ($\text{Cy}_{*}^{\text{R}}$) as the number of hydrogen bonds made to arginine residues by the residue with the most hydrogen bonds to arginine (here, the '*' wildcard means *any residue*). The *residue-specific arginine clawicity* ($\text{Cy}_{i}^{\text{R}}$) is defined as the number of hydrogen bonds made to arginine by each residue, $i$. Thus, for a peptide containing $n$ residues, $\text{Cy}^{\text{R}}$ would contain $n$ values: $Cy_1^R$, $Cy_2^R$, $Cy_3^R$, ..., $Cy_{n-1}^R$, $Cy_n^R$. In the current work, we consider a hydrogen bond to have a donor⋯acceptor distance below 3 Å and a donor–hydrogen–acceptor angle less than 20°. This definition is trivially extended to other residues and may be made smoother by incorporating a definition of hydrogen bonds with non-integer order. For example, the order of a hydrogen bond might smoothly decrease from 1 to 0 as the donor⋯acceptor distance varies from 3 to 4 Å.

As an initial observation, note that representative plots of the arginine clawicity $Cy_*^R$ (Figure 3a) and the site-specific $Cy^R$ (Figure 3b) reveal a rather strong RC when the R2834H $S_{PO_3}2849$ system is simulated under the CHARMM36 force-field.

Include Figure 3 here.

To facilitate communication in this text, we represent $Cy_*^R$ values using histograms as in-line strip charts, e.g. 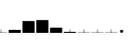. Each bar denotes the frequency of a particular $Cy_*^R$ value across a trajectory, with the leftmost bar representing an $Cy_*^R$ of zero. For example, 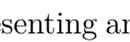 tends to adopt structures of $Cy_*^R$ equal to 1, 2 or 3. Conversely, 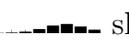 shows a system with a particularly strong RC. Distributions of $Cy_*^R$ values for the last 100 ns of each simulation system are shown in Table 1.

Table 1 NEAR HERE.

We discovered an RC in the conformational states sampled by the $DP_{CTT}$, as shown in Figure 2. Several arginine residues in $DP_{CTT}$ surround $S_{PO_3}2849$ and form numerous hydrogen bonds and ion-pairs. Notably, some of the RCs found in the $DP_{CTT}$ are long-lived structures, such as were those identified by Hamelberg et al.[35] To our knowledge, $DP_{CTT}$ is the largest unstructured peptide wherein an RC has been found.

**Non-phosphorylated $DP_{CTT}$ systems do not form strong claws—** The unmodified (non-phosphorylated) wild-type $DP_{CTT}$ peptide does not adopt a strong RC, as shown in Figures S1a and S2a for the Amber and CHARMM force-fields, respectively. Simulations under PARM99SB show little RC formation (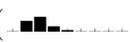), and Figure S1b shows that no residue consistently hydrogen-bonds with any arginine with an $Cy^R$ exceeding unity. The

simulations using CHARMM36 predict slightly higher average $Cy^R_*$, 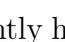, than do those using PARM99SB. Amino acids D2851 and H2871 (the final C′-terminal residue) account for most of the $Cy^R_*$, as shown in Figure S2b.

As mentioned above, a newly-discovered PTM in the $DP_{CTT}$ is asymmetric dimethylation of R2834, yielding $R_{Me_2}2834$.[27] For this modified peptide system, we find a slight increase in the average $Cy^R_*$ when PARM99SB is used, 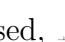. D2851 is the primary contributor to this weak RC (Figure S14b). CHARMM36 predicts a slightly higher $Cy^R_*$ than that seen in the unmodified peptide: 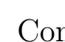. Consistent with the PARM99SB simulation of this system, D2851 is the primary residue creating the RC in the CHARMM36 trajectories (Figure S15b). This particular RC structure does not appear to be dynamically stable: it briefly dissociates 40 ns into the trajectory, and then re-forms at ≈60 ns. This observation suggests that, although a claw can form in this system, the $DP_{CTT}$ would be unlikely to adopt a collapsed RC conformation as a stable, long-lived structure.

The R2834H mutant exhibits low $Cy^R_*$ values under PARM99SB (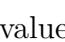), with Figure S8b showing E2804 forming the center of a weak RC. Under CHARMM36, D2851 forms no RC and the overall $Cy^R_*$ is low: 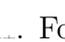. For R2834H simulations under both FFs, the clawicity behavior is similar to that in the unmodified system.

**The behavior of $DP_{CTT}$ is sensitive to force-field**— The backbone dihedral angle distributions for PARM99SB and CHARMM36 are shown in Figure 4. A recent methodological study of an arginine/serine (RS)-rich peptide (unrelated to DP), using several FFs, found that CHARMM36 tends to favor the formation of left-handed helices.[54] We found that $DP_{CTT}$, which also contains an SRR, does not show this trend, at least not on the timescales of our present simulations. Instead, CHARMM36 frequently predicts more $\beta$-strand character (54.5%) than does PARM99SB (40.9%), as indicated in Figure 4. The total helical content (including left-handed helices) is somewhat higher under PARM99SB (23.0%) than it is under CHARMM36 (18.4%).

Include Figure 4 here.

Site R2834H provides a striking example of the differential structural effects of various FFs. In the phosphorylated R2834H system, PARM99SB predicts that H2834 will be essentially entirely buried in the protein, or at least occluded from solvent (Figures S10d and S11d). In contrast, CHARMM36 predicts that this residue will be solvent-exposed, perhaps as a result of the constraints imposed by the strong RC that forms in this system (Figures S12d and S13d). Similarly, in the methylated system, PARM99SB predicts a more buried $R_{Me_2}$ (Figures S16d and S17d) than that predicted by CHARMM36 (Figures S18d and S19d).

**Phosphorylation of S2849 leads to claw formation in the wild-type system**—Trajectories computed under both CHARMM36 and PARM99SB are consistent, inasmuch as the $S_{PO_3}2849$ system (with no other modifications) frequently forms an RC. PARM99SB predicts a substantial shift from the unmodified $Cy_*^R$ profile, 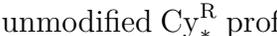. The $S_{PO_3}2849$ system adopts a high $Cy_*^R$, 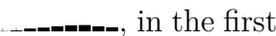, in the first 100 ns of the production run, followed by 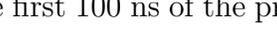 in the next 100 ns (cycle2). Figures S4a and S5a indicate that this system's $DP_{CTT}$'s claw is less stable than that reported for the $(RS)_8$ peptide,[35] and Figure S4a also shows a dramatic re-structuring at ≈70 ns in the production run. CHARMM36 shows a similar trend, moving from the unmodified $Cy_*^R$ (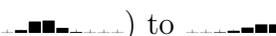) to 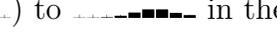 in the first 100 ns, and then 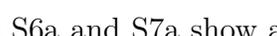 in the second 100 ns (cycle2). Figures S6a and S7a show a more stable RC, akin to that seen previously.[35] For both FFs, the RC that forms is centered around position $S_{PO_3}2849$ (panels (b) in Figures S4 to S7).

For comparative purposes, an additional simulation was performed with the monoprotic phosphate modification, $S_{HPO_3}2849$ (versus the diprotic $S_{PO_3}$), using the Amber FF. This system, under PARM99SB, exhibited essentially no $Cy_*^R$ (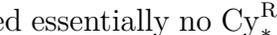). Figure S3b shows that the RC does not form around $S_{HPO_3}2849$ to any appreciable extent; instead, an aspartate residue (D2851) makes occasional structures with $Cy_{D2851}^R$ values of 2, and this tendency diminishes after ≈40 ns.

**Methylation of R2834 weakens the RC**— Simulations of the phosphorylated peptide with PARM99SB show that methylation of R2834, in conjunction with the phosphorylation at S2849, significantly weakens the RC, with ▁▃▅▃▁ in the first 100 ns followed by ▁▃▅▃▁ in the second 100 ns. The second cycle even has slightly lower $Cy_*^R$ than the non-phosphorylated R2834H system. Figures S16b and S17b show that the principal residue involved in the RC is still $S_{PO_3}2849$. The effect predicted by CHARMM36 is more subtle: the $Cy_*^R$ values remain similar to the non-methylated system in terms of their distribution, but Figures S18a and S19a show that the RC is more labile in this system. The sliding-window average (green trace) shows an increased variability compared to the nearly-constant behavior seen in Figures S6a and S7a (compare also the panels (c) [SASA of S2849] in Figures S6, S7, S18 and S19). Again, the CHARMM36 RC is centered on $S_{PO_3}2849$ (Figures S18b and S19b).

**Mutation R2834H may disrupt the RC structure**— For simulation systems containing the R2834H point-mutant as well as phosphorylation at S2849 (i.e., $S_{PO_3}2849$), the two FFs give differing results. Specifically, PARM99SB predicts essentially no change from the non-phosphorylated system in terms of $Cy_*^R$, with ▁▃▅▃▁ for the first 100 ns and ▁▃▅▃▁ for the next 100 ns; Figures S10b and S11b show that the RC stably settles at $S_{PO_3}2849$ by the second half of the 200-ns trajectory. In contrast, the CHARMM36 simulation of this system gives the strongest RCs observed in any of our trajectories (▁▃▅▃▁ for the first 100 ns, followed by ▁▃▅▃▁ for the remaining 100 ns), with the RC forming essentially near the start of the trajectory. The running averages for CHARMM36 (Figures S12a and S13a) reach clawicity values of 6, while no other simulation system ever reaches 5.

## RCs typically exclude solvent

Analysis of the solvent-accessible surface area (SASA) of $S_{PO_3}2849$ can be used to reveal the general solvation features (buried, partially exposed, or fully exposed) of the RC in a 3D structure. The negatively-charged $S_{PO_3}$ residue will be electrostatically attracted to arginine side-chains and, as expected, this is borne out in our observations of $Cy_*^R$ values. In those simulation systems containing $S_{PO_3}2849$ but not exhibiting an RC, one might expect that the phosphate would be solvent-exposed and engaged in hydrogen bonds with water. We find that, for the non-methylated systems, this model works well. In systems containing a strong RC, a plot of the SASA of $S_{PO_3}2849$ shows that the phosphoserine is buried within the protein, as seen by comparing the (a) (RCy) and (c) (SASA) pairs of panels in Figures S4 to S7, S12 and S13, and note the anticorrelation between RCy values and the SASA. In the PARM99SB simulation of the S2849-phosphorylated R2834H mutant, the $Cy_*^R$ was relatively low in the first (Figure S10a) and second (Figure S11a) 100-ns bins, and this agrees with the higher SASA observed for $S_{PO_3}2849$ in Figures S10c and S11c; the negative correlation can be seen here, again, most clearly by comparing the trend in Figure S10a (increasing values) and Figure S10c (decreasing values). The monoprotic system, $S_{HPO_3}2849$ under the Amber FF, similarly shows low $Cy_*^R$ and high SASA values for $S_{HPO_3}2849$ (Figure S3c); the SASA values at this site are quite broadly distributed (Figure S3c, marginal histogram), implying a structurally heterogeneous ensemble of conformational states.

The methylated systems present two deviations from this inverse trend between $Cy_*^R$ and SASA values. Under PARM99SB, $S_{PO_3}2849$ interacts with non-arginine residues on the protein surface, in the methylated system. In Figure S17c, the SASA of $S_{PO_3}2849$ can be seen to jump from a buried state to an exposed state after 40 ns, with no concomitant change in the $Cy^R$ (Figure S17b). $S_{PO_3}2849$ interacts with S2861 and S2835 until 140 ns in the production trajectory, at which point it disengages from these residues while remaining attached to R2838 (until 197 ns). When the methylated system is simulated using CHARMM36, a solvent-exposed RC forms. The phosphate is clearly solvent-exposed, as shown in Figures S18c and S19c, but this system still forms an RC (▁▁▃▄▆▇▆▄▃▁▁).

## Methylation and phosphorylation prime DP$_{\text{CTT}}$ for GSK3 activity

The DP$_{\text{CTT}}$ sequence (Figure 1) contains several potential phosphorylation sites, including consensus sites for the GSK3 kinase. Recent experiments have revealed that DP is phosphorylated in its CTR by GSK3.[27] Thus, we used two simple metrics to assess the ability (not necessarily the propensity) of the DP$_{\text{CTT}}$ to interact with GSK3 throughout the entire MD trajectory: (i) the S$_{\text{PO}_3}$2849⋯⋯S2845 distance, and (ii) the extent of steric clash between the DP$_{\text{CTT}}$ and GSK3 molecules. First, the simple geometric distance between S$_{\text{PO}_3}$2849 and S2845 was measured and compared to the distance between the recognition site and active site in GSK3. This distance was used because S$_{\text{PO}_3}$2849 maps to GSK3's recognition site and S2845 corresponds to the kinase's active site. In the GSK3 crystal structure,[24] this distance is ≈ 12 Å (some variability in this value is expected, as the active site was not occupied by a substrate in this GSK3 structure). As a rudimentary gauge of DP$_{\text{CTT}}$'s ability to bind to GSK3, we suggest that DP$_{\text{CTT}}$ conformations wherein the S$_{\text{PO}_3}$2849⋯⋯S2845 distance is ≈ 12 Å will be more favored to bind to GSK3 as a result of simple geometric matching, without requiring substantial structural rearrangement of the DP$_{\text{CTT}}$.

While DP$_{\text{CTT}}$ systems that are not phosphorylated at S2849 would not be expected (biologically) to interact with GSK3, it is nevertheless informative to consider, as a background distribution, how these distances compare for the non-phosphorylated and phosphorylated systems. We find that the distances in the non-phosphorylated systems show a strong dependence on FF. PARM99SB yields distances that are substantially less than 12 Å for the completely unmodified wild-type system (Figure S1e) and the methylated, non-phosphorylated wild-type system (Figure S14e). The non-phosphorylated R2834H mutant system starts with GSK3-compatible distances, but collapses at ≈ 70 ns to incompatible distances (Figure S8e). In general, the CHARMM36 simulations predict longer distances than PARM99SB, and tend to predict distances that are more compatible with GSK3 binding (see Figures S2e,

S9e and S15e).

Simulations of the phosphorylated DP$_{\text{CTT}}$ systems exhibit good agreement between the distance distributions for PARM99SB and CHARMM36. Our distance parameter consistently lies between ≈12–15 Å, which is compatible with GSK3 binding. The R2834H mutation in the phosphorylated system leads to a slight decrease in the distance under both PARM99SB and CHARMM36 (panels (e) in Figures S10 to S13), compared to that seen in the other two phosphorylated wild-type systems—namely, (i) the phosphorylated (S$_{\text{PO}_3}$2849) system in panels (e) of Figures S4 to S7, and (ii) the phosphorylated & dimethylated systems (S$_{\text{PO}_3}$2849 & R$_{\text{Me}_2}$2834) in panels (e) of Figures S16 to S19.

Our second GSK3-compatibility criterion, described in Figure 5 and in the Methods section, assesses the ability of GSK3 to sterically accommodate various structural states of DP$_{\text{CTT}}$. Specifically, we align (i) the active site of GSK3 with S2845 of DP$_{\text{CTT}}$ (as this is where the next phoshporylation event will occur), and (ii) the recognition site of GSK3 to S$_{\text{PO}_3}$2849 (as this is the landmark in DP$_{\text{CTT}}$ that is recognized). These spatial transformations and geometric constraints effectively reduce the problem to a one-dimensional protein•protein docking exercise, the one degree-of-freedom being rotation about the line defined by constraints (i) and (ii); this construction is schematized in Figure 5. If there exists a rotation wherein GSK3 and DP$_{\text{CTT}}$ can be brought together without substantial steric clash (literally, overlap of atomic van der Waals envelopes), then this suggests that GSK3 can readily bind to that conformation of DP$_{\text{CTT}}$ (or at least that there is no enthalpic barrier to doing so). By this measure, we find that the only phosphorylated DP$_{\text{CTT}}$ systems which exhibit steric compatibility along the trajectory frames are the methylated systems (Figures S17f and S19f). This accommodation is seen with both the PARM99SB and CHARMM36 FFs in the last 50 ns of the production run. Therefore, based on these data we suggest that methylation at R2834, yielding R$_{\text{Me}_2}$2834, 'primes' DP$_{\text{CTT}}$ for processive phosphorylation by biasing its structural ensemble towards conformations that are amenable to GSK3 phosphorylation.

*Include Figure 5 here.*

## The serine-rich region of DP$_{\text{CTT}}$ is not entirely free in solution

Potential interactions between the SRR of the DP$_{\text{CTT}}$ and the rest of the large DP protein (Figure 1) were explored by analyzing pairwise inter-residue distances. As detailed in the Methods section, the full suite of contact maps, shown in panels (g) of Figures S1 to S19, give the mean inter-residue distances (lower triangle), averaged over entire trajectories, while the upper-right triangle gives the minimum inter-residue distance across an entire trajectory.

One may be tempted to view the DP$_{\text{CTT}}$ as a disordered string that thermally fluctuates in solution, but this is not entirely accurate: the dynamical DP$_{\text{CTT}}$ may in fact double back on the plakin repeat domains (as a reminder, see the PRDs in Figure 1). While simulations of larger DP systems, including an entire PRD in addition to the DP$_{\text{CTT}}$, are beyond the scope of this work, the first few residues of our DP$_{\text{CTT}}$ simulation system are from a PRD (the third PRD in Figure 1). Therefore, if the phosphorylated S$_{\text{PO}_3}$2849 samples conformations that bring it near the first few residues of DP$_{\text{CTT}}$, then that suggests that regions within the DP$_{\text{CTT}}$ may interact directly with the PRDs to regulate IF binding (and also that simulations limited to only the SRR might not account for all the factors that govern the structure and dynamics of this region). In all of our simulations, the SRR comes into close spatial proximity to other regions of DP$_{\text{CTT}}$, including the more N′-terminal residues that are part of PRD-C. A possible mechanism by which the DP$_{\text{CTT}}$ can attenuate the overall strength of DP•IF binding may involve a simple binding competition between IFs and DP$_{\text{CTT}}$ for the IF-binding site of the plakin repeat domain; in this model, the precise pattern of PTMs, and therefore the clawicity and dynamics of the DP$_{\text{CTT}}$, would modulate the competitive binding events. When DP$_{\text{CTT}}$ is fully phosphorylated, its strongly negative charge could compete with the (negatively-charged) IFs for the binding groove on the PRD, as suggested by crystallographic studies.[8,10,12]

*Include Figure 6 here.*

## Discussion

**Arginine claws can form in partially phosphorylated systems**— Past work on the arginine claw considered only fully-phosphorylated $(RS)_n$ repeats.[35,38] To our knowledge, our present study provides the first evidence that RCs can form in other systems too. Experimental and computational studies of a 36-residue peptide from myelin basic protein suggested that phosphorylated threonines can confer structure to disordered regions, via electrostatic interactions with basic residues. However, unlike an RC, the interactions in that system did not result in burial of the phosphate group within the protein.[47] Our present simulations, focused on the $DP_{CTT}$, predict that a strong RC can form in protein segments with only half the arginine density of RS-repeat peptides, and even when only a single serine is phosphorylated. Therefore, the RC may be a common, or at least underappreciated, structural element in phosphorylation-based regulation of protein function via molecular switches, even for protein sequences that lack canonical $(RS)_n$ repeat regions.

**Claws are predicted by several force-fields**— The original RC was described as "very stable and, once formed, persist[ing] for the rest of the simulation";[35] that initial study employed only the Amber FF03 parameter set. A subsequent study of another RS-rich peptide found that RCs form under the Amber PARM99SB-ILDN FF.[38] Our simulations of $DP_{CTT}$ show that RCs can form under both PARM99SB and CHARMM36. Nevertheless, the fine details of RC dynamics are sensitive to the FF; for instance, for many of our systems CHARMM36 frequently predicts higher $Cy_*^R$ values than does PARM99SB.

The FF-dependence of our $Cy_*^R$ parameter is substantial, and this may reflect the somewhat unusual chemical nature of RC sequences, versus most protein sequences. In addition to charged moieties buried in a proteinaceous core, arginine···phosphate interactions are characterized by a "covalent-like" stability[83] that may be only inadequately described as point-charges interacting via simple Coulombic electrostatics. An RC was not detected in recent NMR experiments with another RNA splicing-related, serine/arginine-rich system;[37]

however, the structural ensembles reported in that work were derived via an approach (a 'sub-ensemble selection procedure' against the NMR data) differing from the simple, naive equilibrium MD simulations reported here and elsewhere,[35,38] and the trajectories in that work sampled shorter (≈50-ns) timescales. In short, it remains to be established if RCs occur in solution, and under what conditions. A recent crystal structure has shown that (solvent-exposed) RCs can form in the RNA splicing factor SF1.[39] In that system, the RC acts as a secondary structural element in an otherwise disordered region; notably, electron density could be detected for residues immediately upstream of the phosphoserine, but only in the phosphorylated, not the non-phosphorylated, system.[39]

**Methylation in the $DP_{CTT}$ may promote GSK3 binding**— From the simulations presented here, we suggest that the $R_{Me_2}2834$ and $S_{PO_3}2849$ PTMs are required for productive DP⋯GSK3 interactions. This claim is based upon three lines of evidence. First, our modified $DP_{CTT}$ systems were found to present the phosphate group on the surface, rather than buried within the protein. This surface exposure did not occur in phosphorylated systems with unmodified R2834, suggesting that methylation is coupled to the dynamics of $S_{PO_3}2849$ accessibility. As the processive kinase GSK3 recognizes proteins already containing a phosphate, exposure of $S_{PO_3}2849$ may facilitate GSK3 binding. Second, we find that in some trajectories the $S_{PO_3}2849$⋯⋯S2845 distance closely matches the distance between the active site and substrate recognition site of GSK3. Upon GSK3 binding to $S_{PO_3}2849$, S2845 can reach the active site of GSK3 without DP having to undergo conformational changes. Third, the steric clash (Figure 5) between DP and GSK3, computed along entire trajectories, is far lower in systems containing $R_{Me_2}2845$ than in those without this PTM. The degree to which DP must deform to bind to GSK3 is therefore much lower, increasing the probability that contact between DP and GSK3 leads to the addition of a phosphate at S2845; that is, PTMs may help 'pre-structure' the DP substrate in a binding-competent state, thereby decreasing the entropic cost associated with forming a DP•GSK3 complex. In our mechanistic model for GSK3 regulation, $DP_{CTT}$ essentially self-regulates its processive phosphorylation by GSK3; DP$_{\text{CTT}}$ achieves this by sampling conformational states that vary in their suitability as substrates for GSK3.

**The serine-rich region of DP$_{\text{CTT}}$ contacts other parts of DP**— Past studies of RCs have examined short, (RS)$_n$–containing peptides in isolation. The serine-rich region of DP$_{\text{CTT}}$ is not well-described by these past models, as we have shown that the SRR can interact with other regions of DP. In particular, the SRR can contact residues that have been resolved in a crystal structure of a plakin repeat domain.[8] The charge-complementarity between a fully-phosphorylated SRR in DP$_{\text{CTT}}$ and the positively-charged IF-binding groove on a PRD,[8] combined with the tendency for DP$_{\text{CTT}}$ to explore the surface of DP, suggests that a simple competition for PRD binding sites may account for the cellular effects of DP$_{\text{CTT}}$ phosphorylation. That the DP$_{\text{CTT}}$ is covalently linked to the upstream PRDs (Figure 1) implies a high local density of negative charge, and this could compete with the negatively-charged IFs to cause DP to detach from the IF network; examination of the ionic strength-dependence of this process would be telling. Finally, note that in our mechanistic model any structural role for arginine claw conformational dynamics (apart from its role in GSK3 processive phosphorylation) would require a further series of simulations, ideally including as many structured PRD regions as possible.

## Conclusion

Recent experiments have revealed that desmoplakin's activity is regulated by PTMs in its presumably-disordered C-terminal tail. Using MD simulations, we have elucidated the structural effects of three modifications in the 70-residue DP$_{\text{CTT}}$ region: phosphorylation of S2849, methylation of R2834, and mutagenesis of R2834 to histidine. Our simulations indicate that an RC can form in some of the phosphorylated systems, sequestering the phosphate within the protein. To our knowledge, DP$_{\text{CTT}}$ is the largest system that has been shown to form an RC by MD simulation. Our findings build on past studies of RC formation in SR repeats, and are corroborated by recent crystallographic results for other SR systems. Upon methylation of R2834, the $S_{PO_3}2849$ phosphosite becomes solvent-exposed, which may enhance its detection by the cognate kinase GSK3. Methylation of R2834 has the further effect of biasing the structural ensemble towards conformations that are sterically compatible as substrates for GSK3.

We also find that the $DP_{CTT}$'s SRR is not isolated from the rest of DP, suggesting that studies of short peptides excised from larger systems may miss some of the interactions that define the conformational ensemble of such regions. This point is illustrated by the effects of R2834 methylation: The position of the RC, and the overall conformation of the $DP_{CTT}$, are affected by this seemingly minor chemical modification, many residues away from the site of phosphorylation. The common self-contact in $DP_{CTT}$, seen in contact maps for all our simulated systems, suggests that a regulatory mechanism of DP⋯IF adhesion may be a simple binding competition between $DP_{CTT}$ and the IFs for the positively-charged groove on plakin repeat domains.

By elucidating the roles and linkages between protein conformational dynamics, PTMs, and claw-like structural elements, our simulations of the C-terminal region of human desmoplakin synthesize several strands of evidence and shed light on the underlying molecular mechanism of DP⋯IF interactions, including the riddle of strong/weak interactions with the IF network. We predict that RCs can form when S2849 is phosphorylated, and that methylation of the disease-associated site R2834 promotes processive phosphorylation by GSK3. Our data also are consistent with $DP_{CTT}$ binding to a PRD, thus providing a simple, atomically-detailed competition mechanism for the regulation of DP•IF adhesion.

## Associated Content

**Supporting Information** — Detailed results of the analysis suite described in the Methods section of the main text, as applied to each of our simulation systems. Each trajectory has been analyzed in terms of (a) $\text{Cy}^\text{R}_*$, (b) $\text{Cy}^\text{R}$, (c) SASA of S2849, (d) SASA of R2834, (e) the S2849⋯⋯R2834 distance, (f) GSK3 steric clash scores, (g) inter-residue contact maps, and (h) the distribution of peptide backbone torsion angles $(\phi, \psi)$.

# Acknowledgments

We thank D. Hunt & L. Zhang (UVA), as well as K. Green & L. Albrecht (Northwestern), for helpful discussions. We thank A. Dejaegere (IGBMC) for providing force-field parameters for $\text{R}_{\text{Me}_2}$, and D. Hamelberg (Georgia State) for providing the coordinates used to generate Figure 2d. K. Holcomb & A. Munro (UVA) are thanked for exceptional computer support; UVA's Advanced Research Computing Services and Information Technology Services provided computational resources and technical support that contributed to the results reported herein. Portions of this work were supported by UVA, the Jeffress Memorial Trust (J-971), and NSF CAREER award MCB-1350957.

**Figure Captions**, McAnany & Mura, 2016

**Figure 1: Desmoplakin in context: Architecture of the desmosome.** Key components of the desmosome are diagrammed here, with the length of each rectangular protein schematic corresponding to the number of amino acids (scale bar, lower-right). Dsg and Dsc are the transmembrane cadherin proteins desmoglein-1 and desmocollin, respectively; plakoglobin (PG) and plakophilin (PKP) are adapter proteins that bind the N′-terminal region of desmoplakin to the cadherin cytoplasmic regions (CCR) of Dsg and Dsc, as indicated. The PG crystal structure is inset,[85] as are the structures of PKP[84] and two PRDs. DP is shown in the middle, and regions of known structure are inset.[8,10] Crystal structures are drawn as ribbon diagrams, with the color graded from the N′ (blue) to C′ (red) terminus. The locations of the R2834 and S2849 modifications in the $DP_{CTT}$ sequence are marked. Our various DP simulation systems included: (i) the unmodified, wild-type sequence of $DP_{CTT}$, (ii) the wild-type sequence phosphorylated at S2849, (iii) the wild-type sequence methylated at R2834, (iv) the wild-type sequence phosphorylated at S2849 and methylated at R2834, (v) the R2834H mutant, and (vi) the R2834H mutant phosphorylated at S2849. All phosphate PTMs were diprotic, with the exception of one test system containing a monoprotic phosphoserine. (See also Table 1.)

**Figure 2: A recognizable RC in the $DP_{CTT}$.** All simulation systems started in an extended backbone conformation (with bends at each proline residue), as exemplified in (a). After 10 ns of implicit-solvent simulation under CHARMM36, the R2834H S2849$S_{PO_3}$ system can be seen to have collapsed and formed an RC (b). A sample RC, at 95 ns for the R2834H S2849$S_{PO_3}$ system under CHARMM36, is shown in (c). The gray surface surrounds all residues other than arginines (shown as bonds), and the $S_{PO_3}$2849 phosphosite is explicitly shown (ball-and-stick). Hydrogen bonds are represented as dashed orange lines. For reference, an RC previously identified in an unrelated system[35] is shown in (d), rendered similarly as in (c); the coordination geometry of arginines and a phosphoserine is similar to that seen for the $DP_{CTT}$ in (c). The structural stability of an RC is demonstrated in (e) by overlaying multiple frames from the simulation trajectory of (c). The regions near the RC remain stable for the duration of the simulation, with the chelating arginines (shown as bonds) moving very little relative to $S_{PO_3}$2849 (ball-and-stick, only oxygens can be seen). The rest of the $DP_{CTT}$ backbone (thin ribbons) does not adopt a single, stable structure.

**Figure 3: Representative results from the analysis pipeline, showing a strong RC.** The $Cy^R_*$ for the first 100 ns of the R2834H S2849$S_{PO_3}$ simulation under CHARMM36 is shown in (a), demonstrating the appearance of a strong claw (large, persistent clawicity value). Blue points show the arginine clawicity, $Cy^R_*$, defined as the number of hydrogen bonds made to arginines by the residue with the most hydrogen bonds to arginine; the green line is a 1-ns running average. The marginal distribution on the right is a histogram of piled-up $Cy^R_*$ values, ranging from 0 to 8: 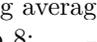. The residue-specific $Cy^R$ (b) shows that $Cy^R_{S2849}$ frequently exceeds 6, and that only D2851 (immediately to the right of the dark strip) makes any other substantial contribution to this system's $Cy^R_*$.

**Figure 4: Backbone conformations across all simulations.** To gauge the frequency of any unusual (non-canonical) secondary structures, Ramachandran plots are shown for all of our (a) CHARMM36 and (b) PARM99SB data, compiled across all trajectories for all simulation systems. Regions corresponding to canonical secondary structural elements are indicated as countour lines (see also the Methods section), and the color-scale is graded by the log-likelihood of a particular $(\phi,\psi)$ conformation. In contrast to a recent report,[54] we find that $DP_{CTT}$ does not show a preference for left-handed helices under CHARMM36; those recent simulations of a non-phosphorylated SRR, unrelated to our DP systems, found that CHARMM36 predicts that over 40% of the residues in the simulated peptide are in left-handed helices, even though only $\approx 6\%$ of the residues in a reference set of known protein structures exhibit such a structure. Our CHARMM36 trajectory data do not indicate that left-handed helices are a problem, at least in our simulation systems. Combining all frames of every simulation, CHARMM36 predicts 9% left-handed helix, while PARM99SB predicts 7%.

**Figure 5: A method to evaluate potential DP⋯GSK3 geometric complementarity via one-degree-of-freedom docking.** We begin with two molecules to be docked, as schematized in (a). Say we know that the red atom on the yellow molecule aligns with the red atom on the blue molecule when the molecules interact. (In this two-dimensional case, only one atom is needed per molecule; in three dimensions, two atoms per molecule are needed to define an axis of rotation.) In step (b), the molecules have been aligned, in arbitrary angular orientation, based only on the positions of their red atoms. Next, the yellow molecule is rotated about the axis defined by the red atoms (c). At each step in the full rotational sweep (1° increments in our implementation), the *clash score* is computed as the number of pairs of atoms (one from DP, one from GSK3) with an interatomic distance less than 2 Å. The best pose (d) is taken as the one with the minimal clash score. This procedure is repeated for each frame along the trajectory. In a realistic example (e), two atoms (red, orange) from GSK3 are used to perform the alignment. The outward-facing oxygen of the phosphate (orange sphere) defines the recognition site, while the active site is defined by the side-chain oxygen of S261 (red sphere) of chain B (gray ribbon). The protein atoms from chain A (dark surface) were used to calculate the clash. The atoms in DP that were used to perform the alignment are highlighted in (f). This frame, from 173.48-ns in the production trajectory of $R_{Me_2}2834\ S_{PO_3}2849$ (CHARMM36), has a very low clash score of 18; all non-hydrogen atoms from DP (dark surface) were used to calculate the clash. The side-chain oxygen of S2845 (red) will be phosphorylated by GSK3 only if S2849 (orange) is phosphorylated. Note that the three terminal oxygens of the phosphate are still engaged in an RC (hydrogen bonds in orange).

**Figure 6: S2849 in close proximity to the PRD.** This frame, from 71-ns in the R2834H, $S2849S_{PO_3}$ simulation under PARM99SB, exemplifies the contacts made between S2849 and residues that are part of the last plakin repeat domain (PRD C) in DP. Residues 2802–2805 are shown as van der Waals spheres on the left, and $S2849S_{PO_3}$ is shown as vdW spheres in the center. These close contacts suggest that the $DP_{CTT}$ can directly interact with the PRDs.

**Table 1: Simulation systems and their $Cy_*^R$ histograms.** $Cy_*^R$ values from the last 100 ns of each simulation are presented as histograms, where the intensity in a particular bin represents the frequency that the system had the corresponding $Cy_*^R$ value. As an example, the bin numbers are explicitly shown in ![hist]₀₁₂₃₄₅₆₇₈, which represents a simulation that frequently displayed $Cy_*^R$ values of 2, 3, and 4 (highest peaks in the histogram). CHARMM36 was consistently found to predict higher $Cy_*^R$ values than PARM99SB; in terms of clawicity, CHARMM36 also predicts a stronger response to phosphorylation.

*to adjacent cell*

*extracellular space*

*desmosomal cadherin (cell⋯cell′)*

*plasma membrane*

CCR | Dsg
PG
**Desmoplakin (DP)**
IF–binding region (cytoskeletal anchor)
PRD A | PRD B | PRD C | CTT
central coiled-coil region (mediates dimerization)

$_{2802}$LLEAASVSSKGLPSPYNMSSAPGSRSGSRSGSRSGSRSGSRSGSRRGSFDATGNSSYSYSYSFSSSSIGH

2834 → $R_{Me_2}$ → H

2849 → $S_{PO_3}$

2871

PKP

CCR | Dsc

*outer dense plaque* (ODP)

simulation system

*inner dense plaque* (IDP)

200   400
number amino acids

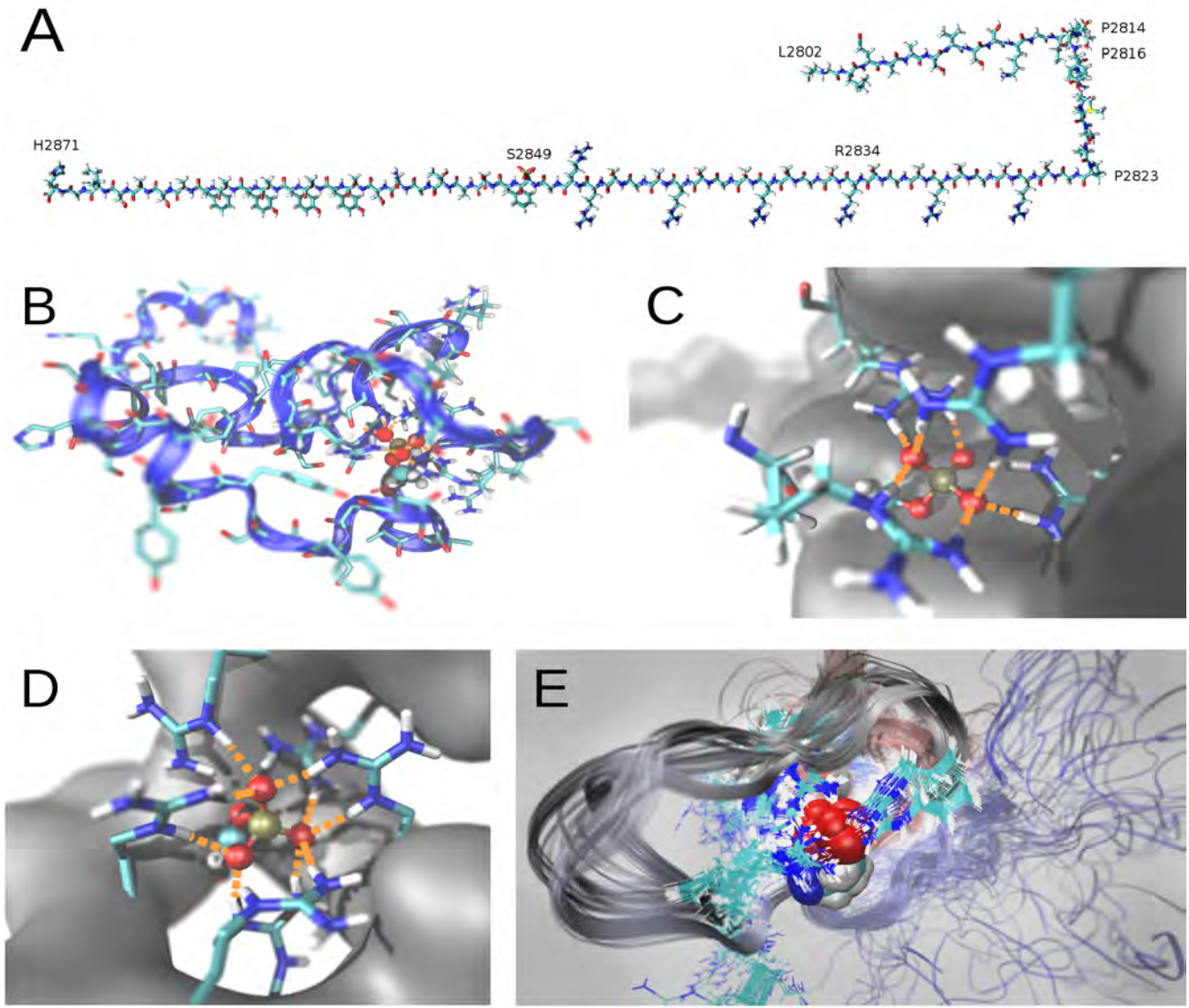

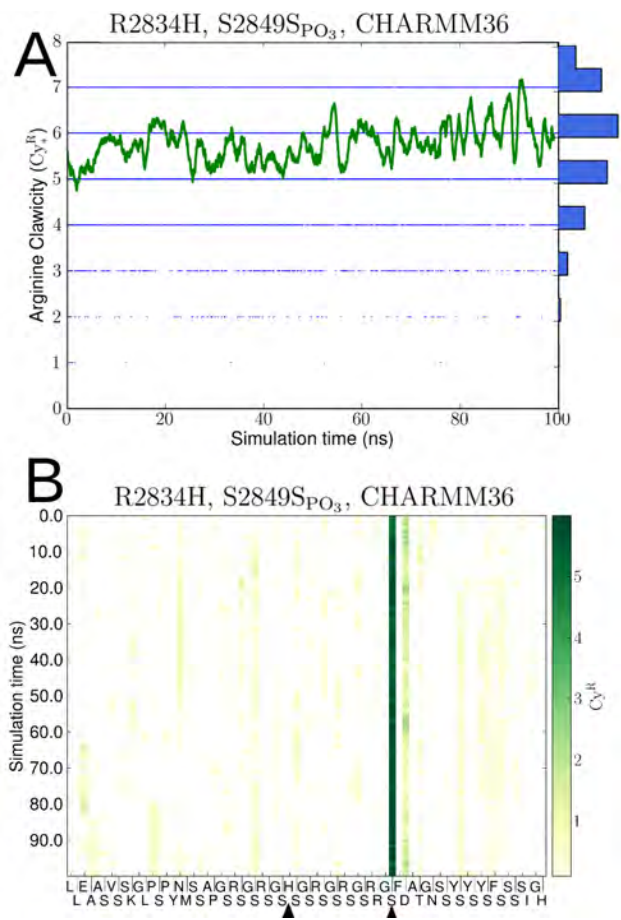

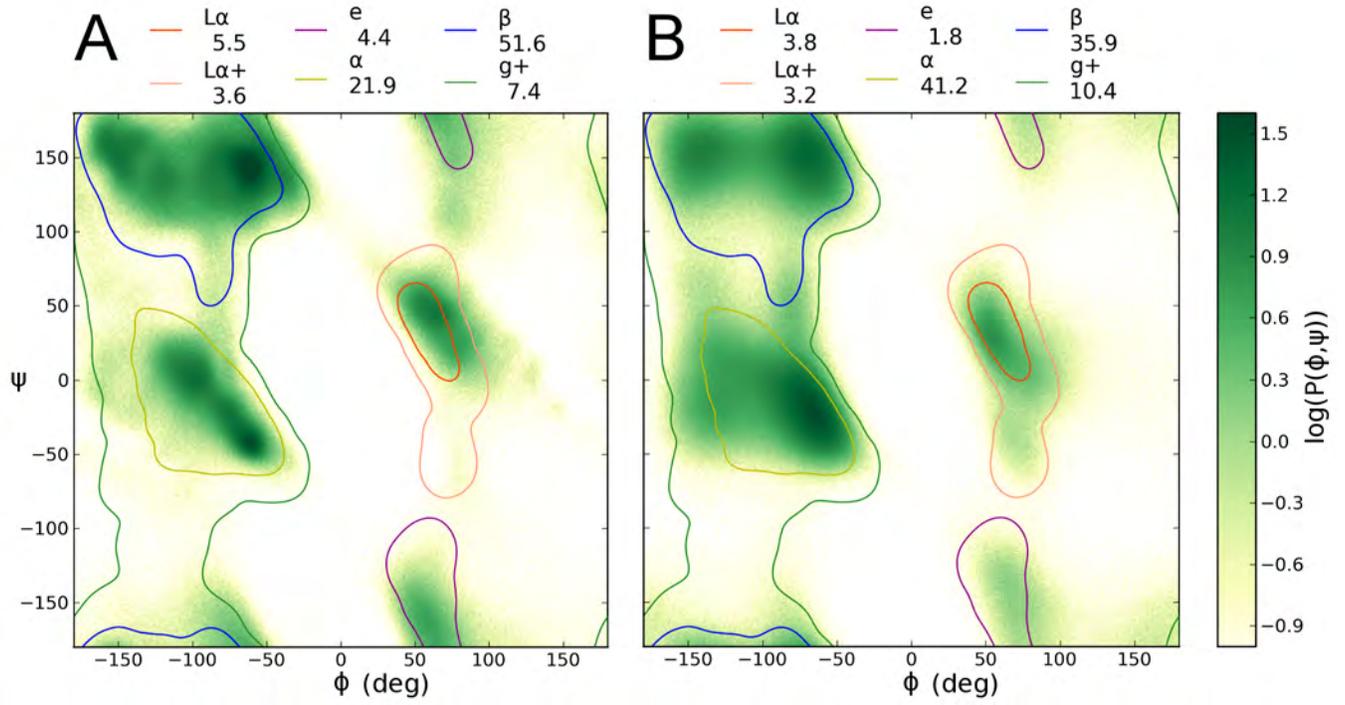

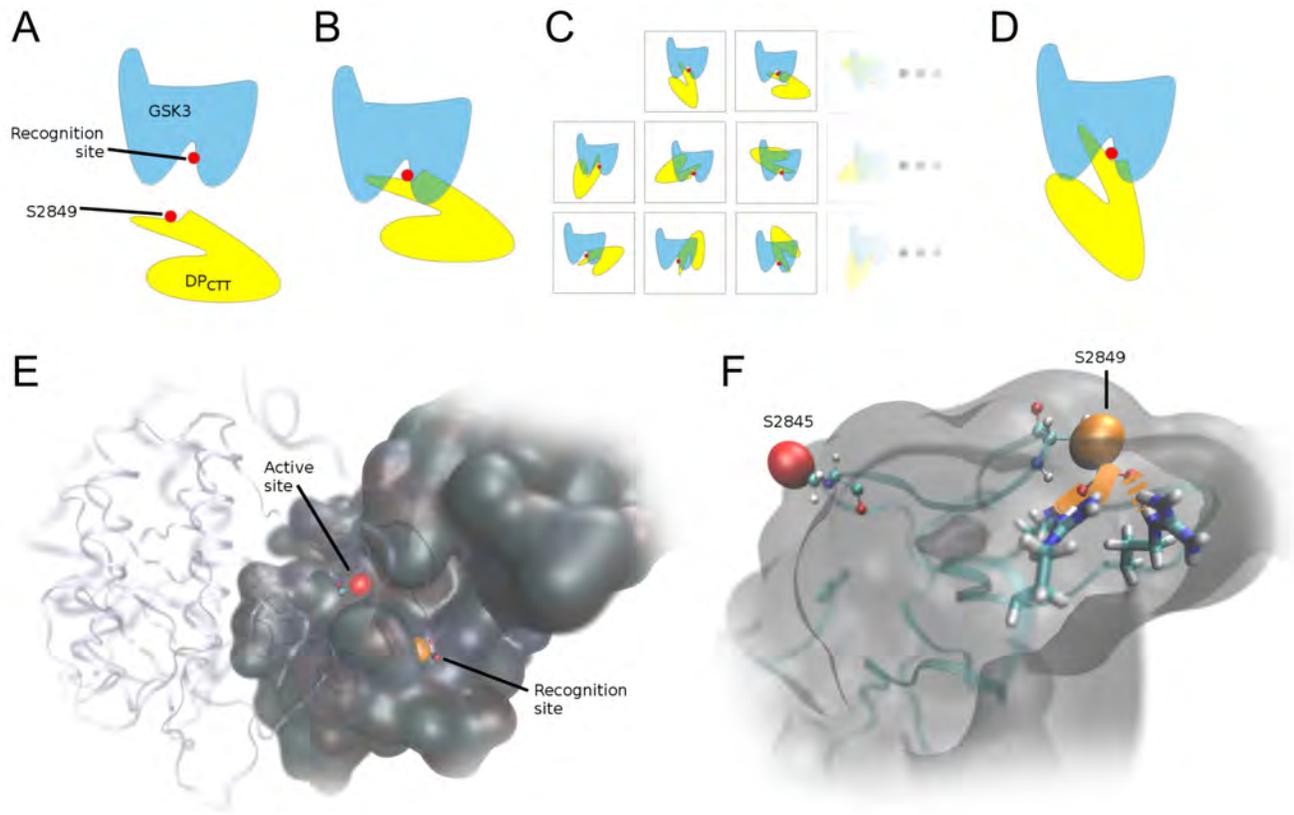

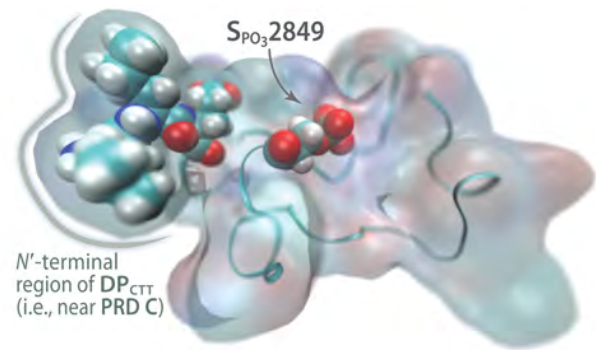

McAnany and Mura (2016), **Table 1**

| Simulation system | Duration (per FF) | Force-field | |
|---|---|---|---|
| | | PARM99SB | CHARMM36 |
| Wild-type, unmodified | 100 ns | | |
| S2849S$_{PO_3}$ | 200 ns | | |
| R2834H | 100 ns | | |
| R2834H and S2849S$_{PO_3}$ | 200 ns | | |
| R2834R$_{Me_2}$ | 100 ns | | |
| R2834R$_{Me_2}$ and S2849S$_{PO_3}$ | 200 ns | | |
| S2849S$_{HPO_3}$ (PARM99SB only) | 100 ns | | – |

# Supporting Information for:

# Claws, Disorder, and Conformational Dynamics

# of the C-terminal Region of Human Desmoplakin

Charles E. McAnany and Cameron Mura[*]

*Department of Chemistry, University of Virginia, Charlottesville, VA 22904, USA*

E-mail: cmura@muralab.org

---

[*]To whom correspondence should be addressed

**Overview**

This document provides the detailed results of the analysis suite described in the Methods section of the main text, as applied to each of our simulation systems. Each trajectory has been analyzed in terms of (a) $Cy^R_*$, (b) $Cy^R$, (c) SASA of S2849, (d) SASA of R2834, (e) S2849·····R2834 distance, (f) GSK3 steric clash, (g) interresidue contact, (h) and backbone dihedral angles.

# List of Figures

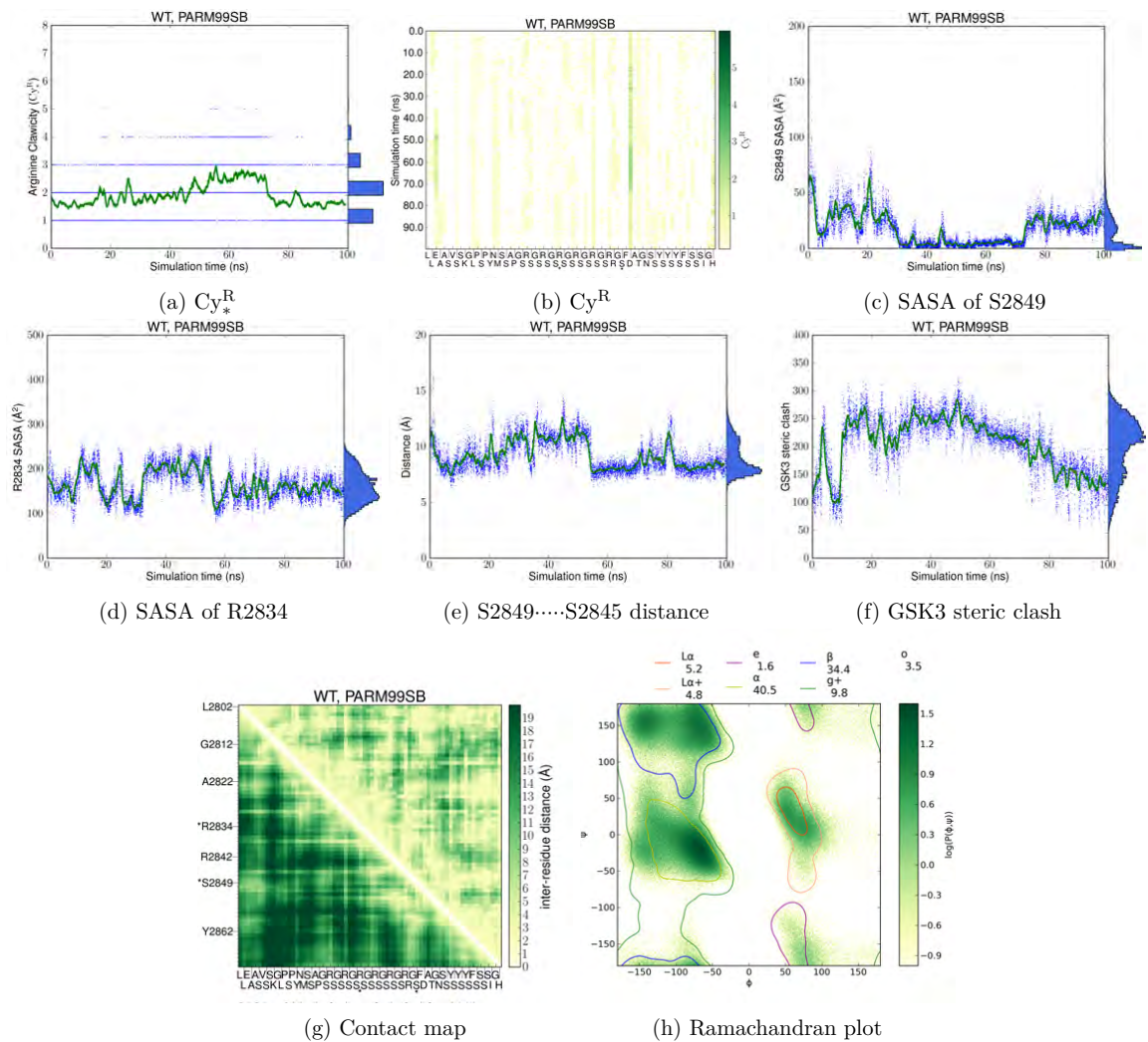

(a) $Cy^R_*$  (b) $Cy^R$  (c) SASA of S2849

(d) SASA of R2834  (e) S2849⋯S2845 distance  (f) GSK3 steric clash

(g) Contact map  (h) Ramachandran plot

Figure S1: Behavior of **WT_PARM99SB**: Time-series plots mark each observation as a blue point and contain a 1-ns running average as a green trace, and the marginal distributions are shown on the right axis, in each of panels a, c, d, e, and f.

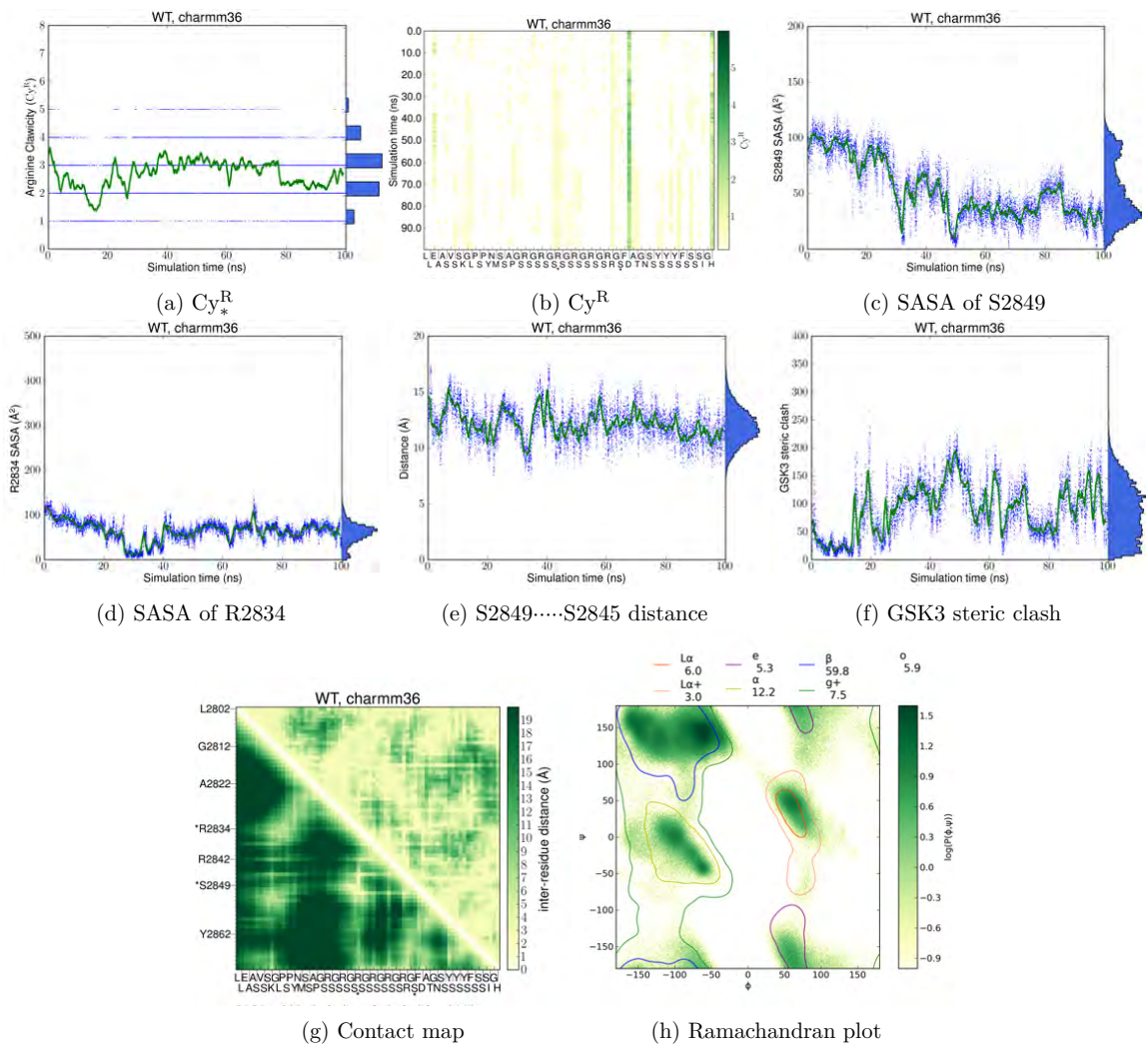

Figure S2: Behavior of **WT_CHARMM36**: Time-series plots mark each observation as a blue point and contain a 1-ns running average as a green trace, and the marginal distributions are shown on the right axis, in each of panels a, c, d, e, and f.

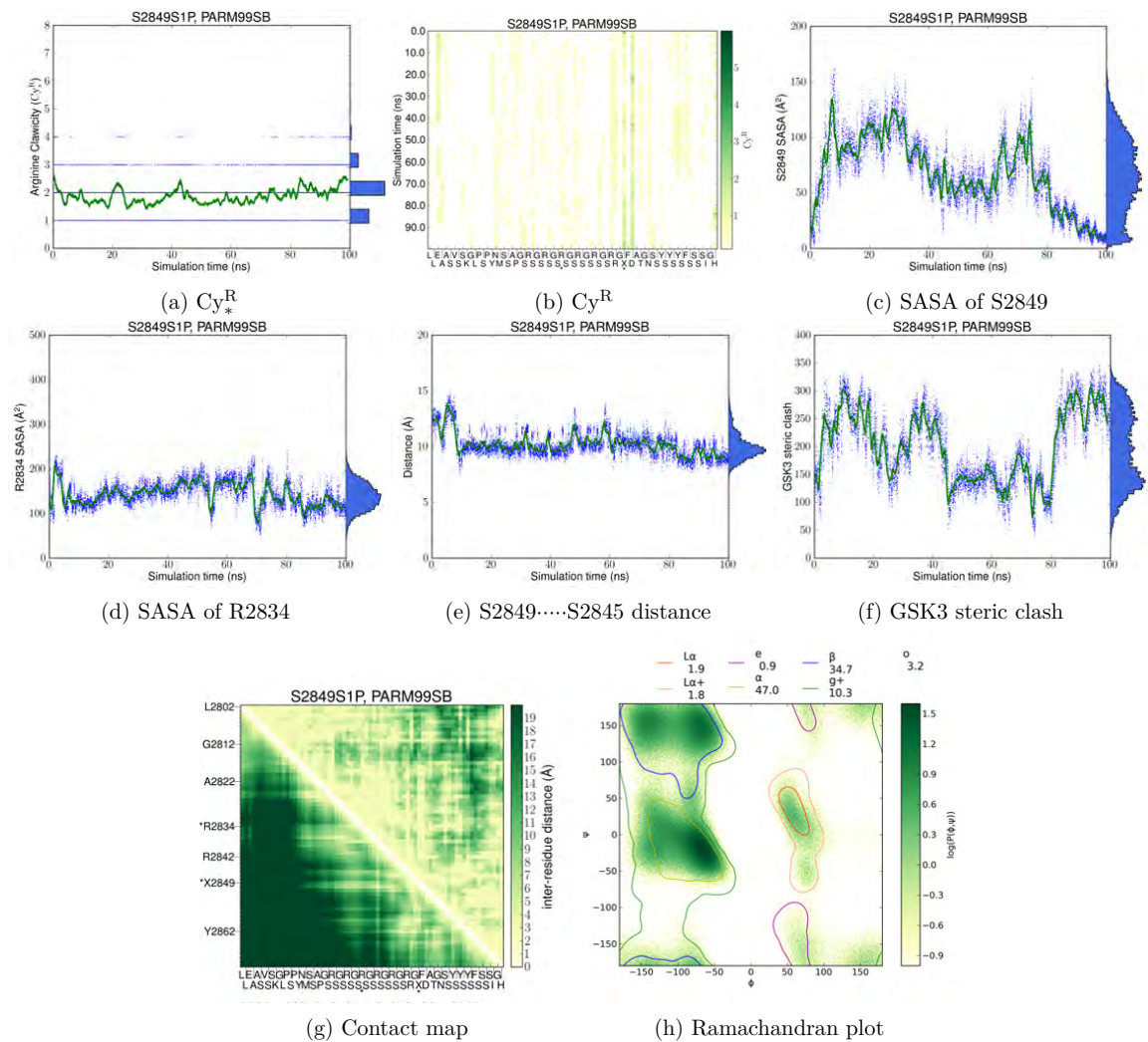

(a) $Cy^R_*$

(b) $Cy^R$

(c) SASA of S2849

(d) SASA of R2834

(e) S2849⋯S2845 distance

(f) GSK3 steric clash

(g) Contact map

(h) Ramachandran plot

Figure S3: Behavior of **S2849S1P_PARM99SB**: Time-series plots mark each observation as a blue point and contain a 1-ns running average as a green trace, and the marginal distributions are shown on the right axis, in each of panels a, c, d, e, and f.

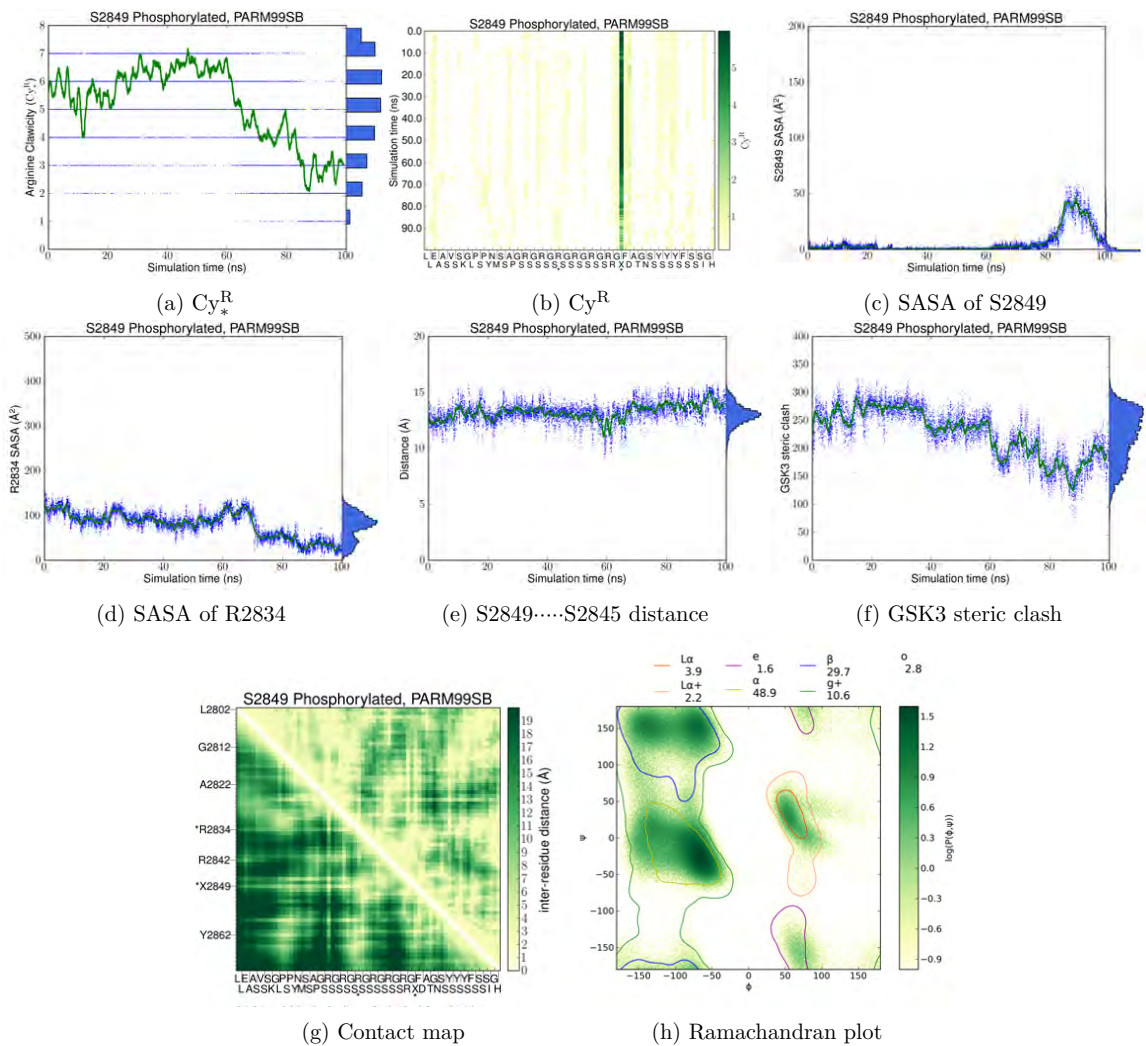

(a) $Cy^R_*$
(b) $Cy^R$
(c) SASA of S2849
(d) SASA of R2834
(e) S2849·····S2845 distance
(f) GSK3 steric clash
(g) Contact map
(h) Ramachandran plot

Figure S4: Behavior of **S2849S2P_PARM99SB**: Time-series plots mark each observation as a blue point and contain a 1-ns running average as a green trace, and the marginal distributions are shown on the right axis, in each of panels a, c, d, e, and f.

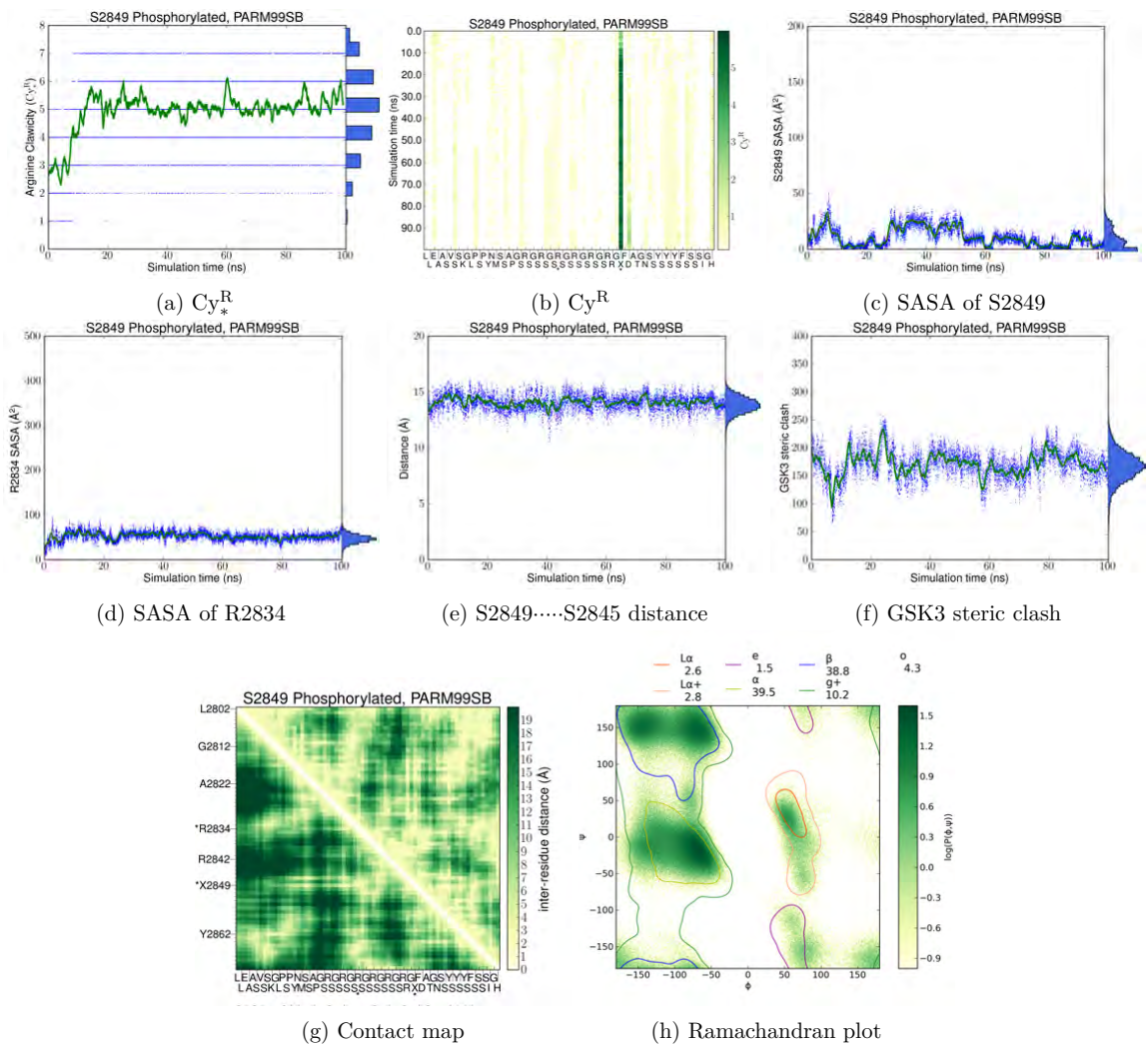

Figure S5: Behavior of **S2849S2P_PARM99SB_cycle2**: Time-series plots mark each observation as a blue point and contain a 1-ns running average as a green trace, and the marginal distributions are shown on the right axis, in each of panels a, c, d, e, and f.

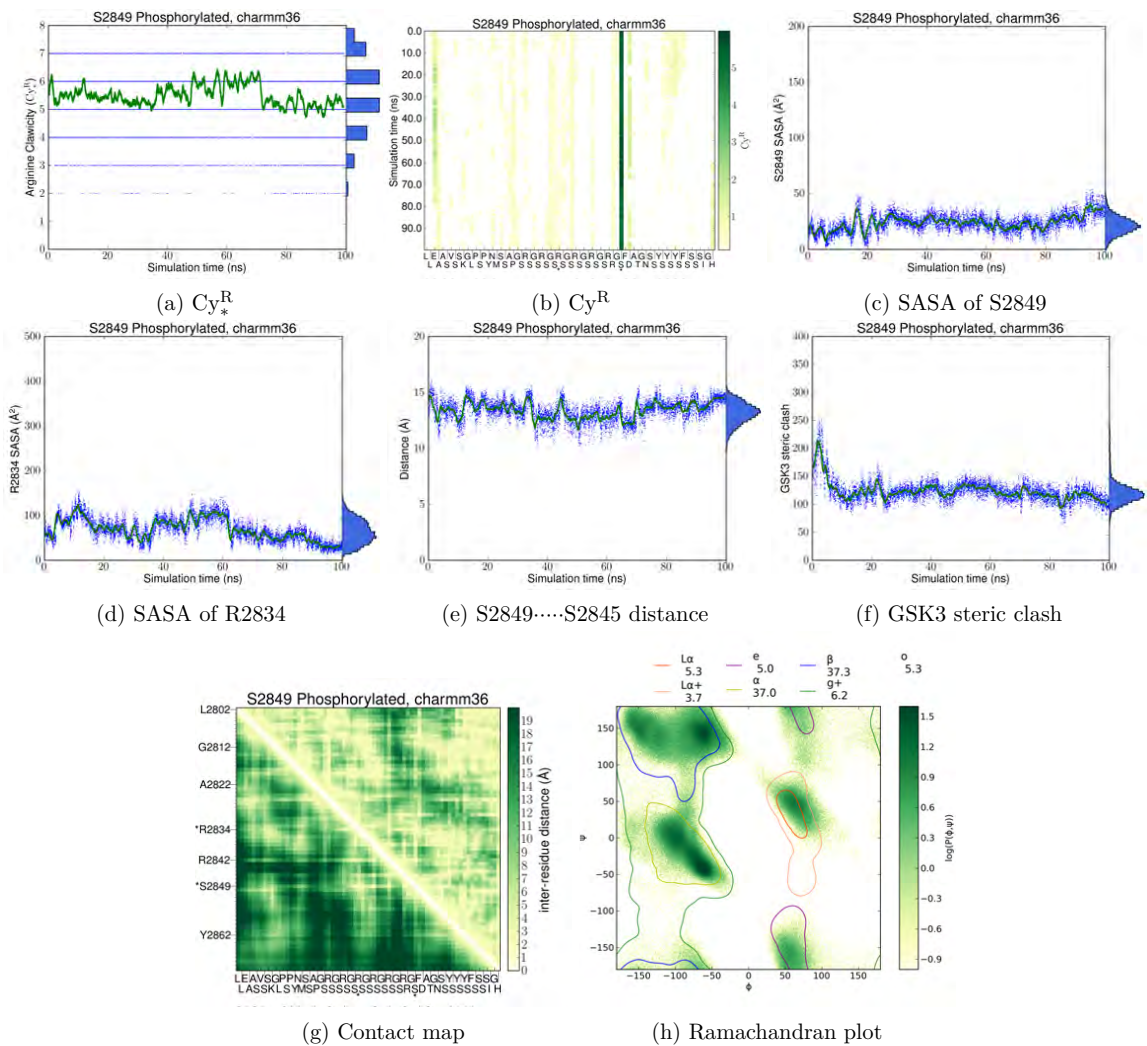

(a) $Cy^R_*$

(b) $Cy^R$

(c) SASA of S2849

(d) SASA of R2834

(e) S2849⋯S2845 distance

(f) GSK3 steric clash

(g) Contact map

(h) Ramachandran plot

Figure S6: Behavior of **S2849S2P_CHARMM36**: Time-series plots mark each observation as a blue point and contain a 1-ns running average as a green trace, and the marginal distributions are shown on the right axis, in each of panels a, c, d, e, and f.

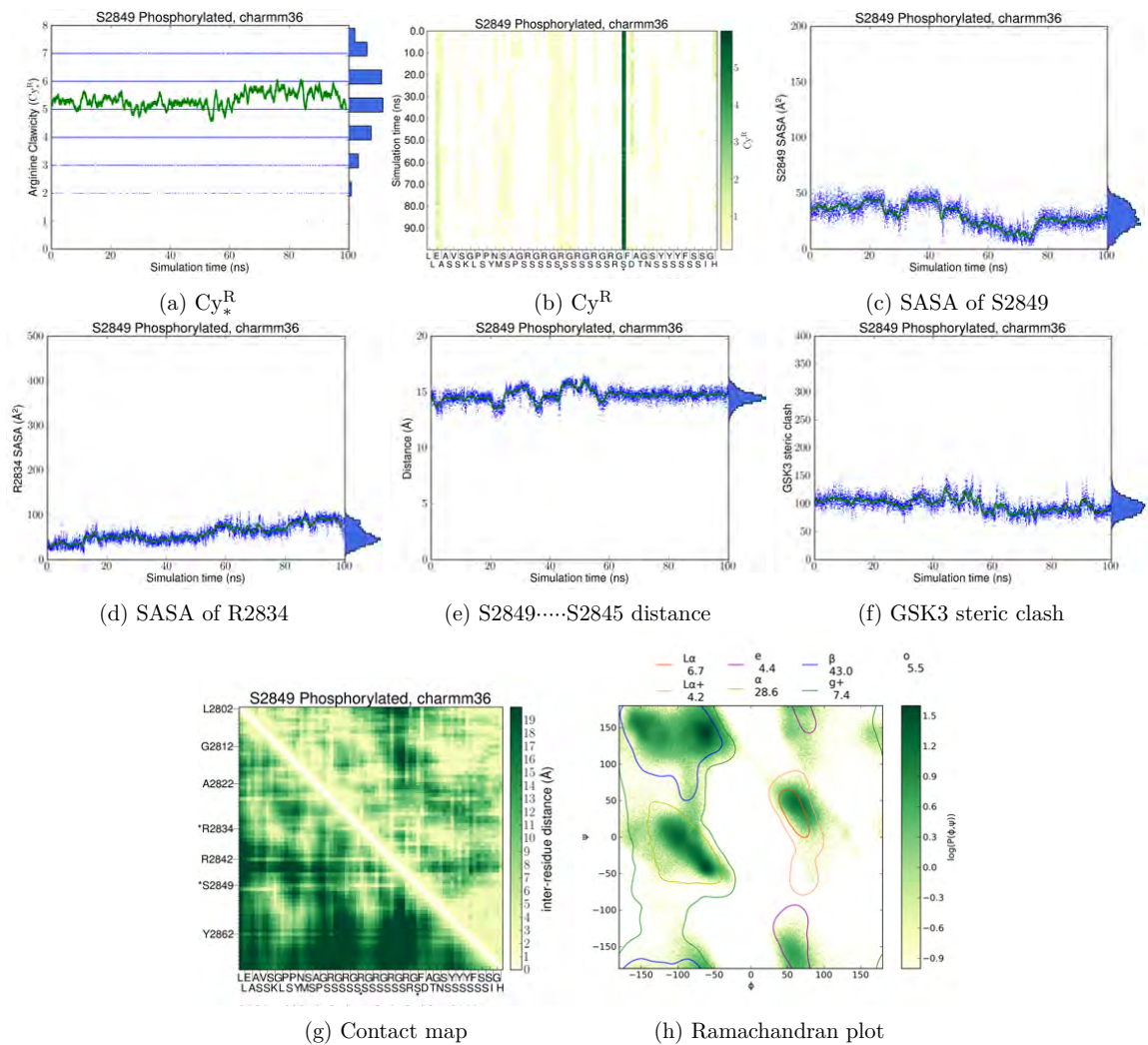

(a) $Cy_*^R$    (b) $Cy^R$    (c) SASA of S2849

(d) SASA of R2834    (e) S2849⋯S2845 distance    (f) GSK3 steric clash

(g) Contact map    (h) Ramachandran plot

Figure S7: Behavior of **S2849S2P_CHARMM36_cycle2**: Time-series plots mark each observation as a blue point and contain a 1-ns running average as a green trace, and the marginal distributions are shown on the right axis, in each of panels a, c, d, e, and f.

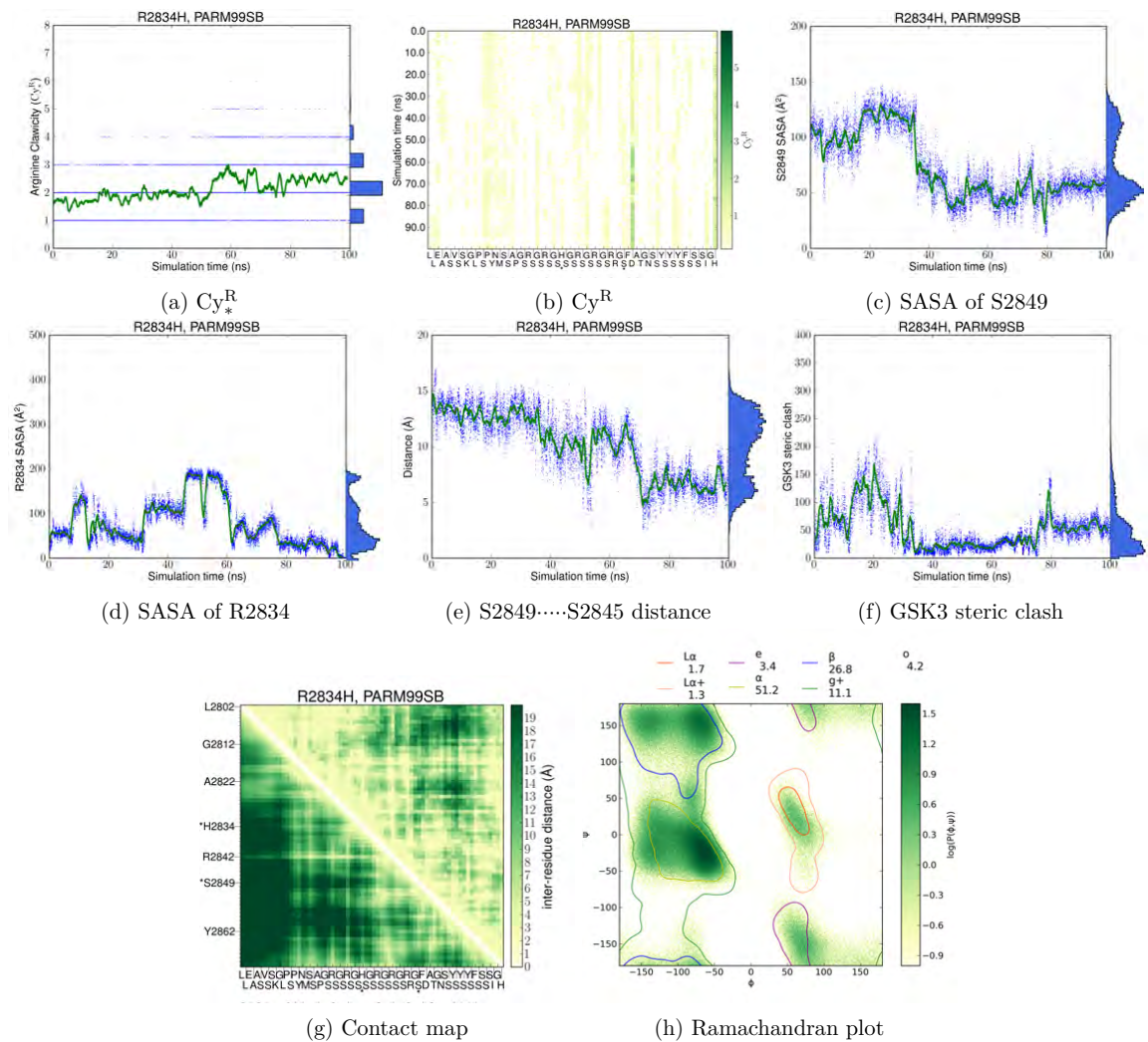

Figure S8: Behavior of **R2834H_PARM99SB**: Time-series plots mark each observation as a blue point and contain a 1-ns running average as a green trace, and the marginal distributions are shown on the right axis, in each of panels a, c, d, e, and f.

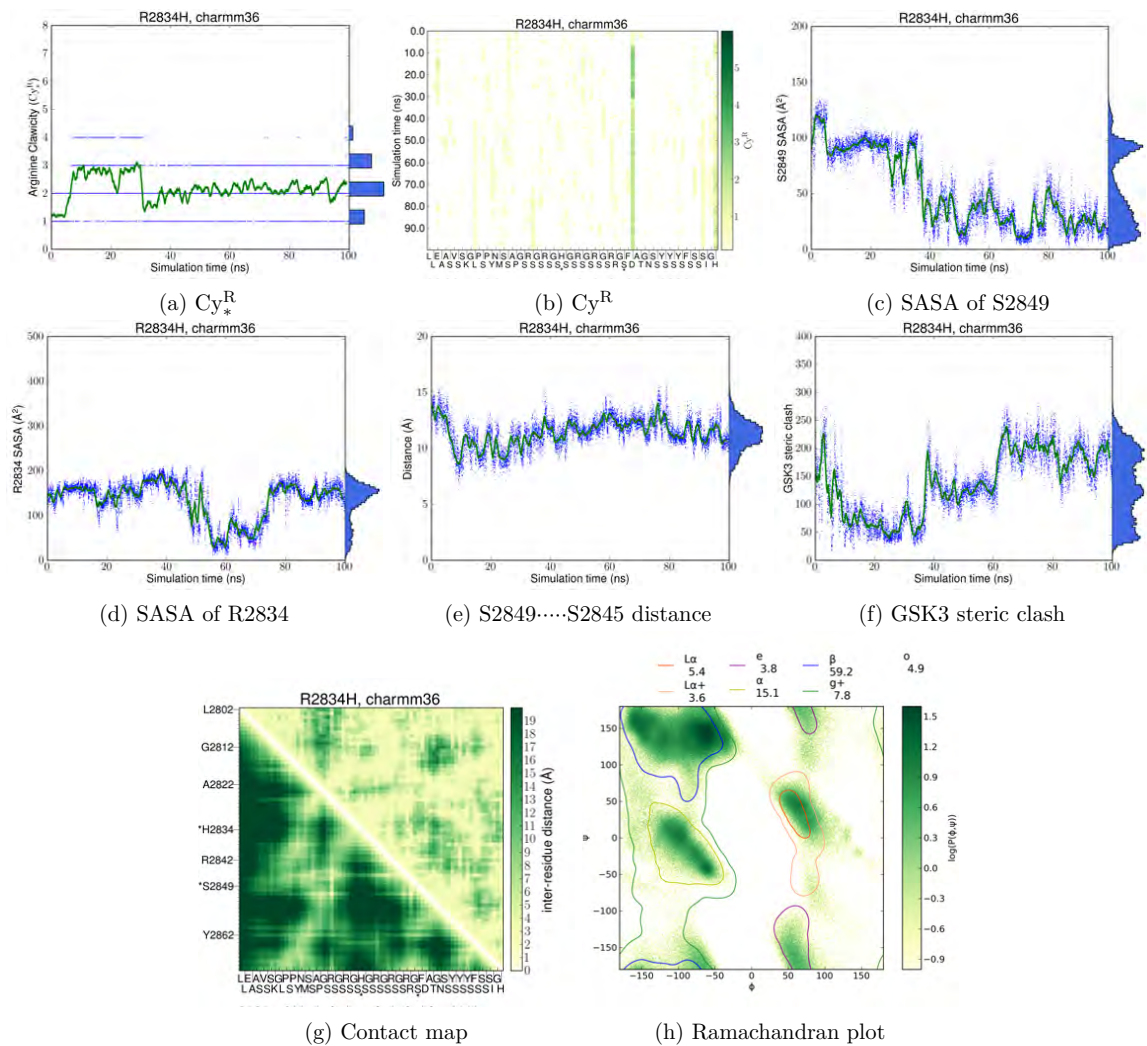

(a) $Cy_*^R$    (b) $Cy^R$    (c) SASA of S2849

(d) SASA of R2834    (e) S2849⋯S2845 distance    (f) GSK3 steric clash

(g) Contact map    (h) Ramachandran plot

Figure S9: Behavior of **R2834H_CHARMM36**: Time-series plots mark each observation as a blue point and contain a 1-ns running average as a green trace, and the marginal distributions are shown on the right axis, in each of panels a, c, d, e, and f.

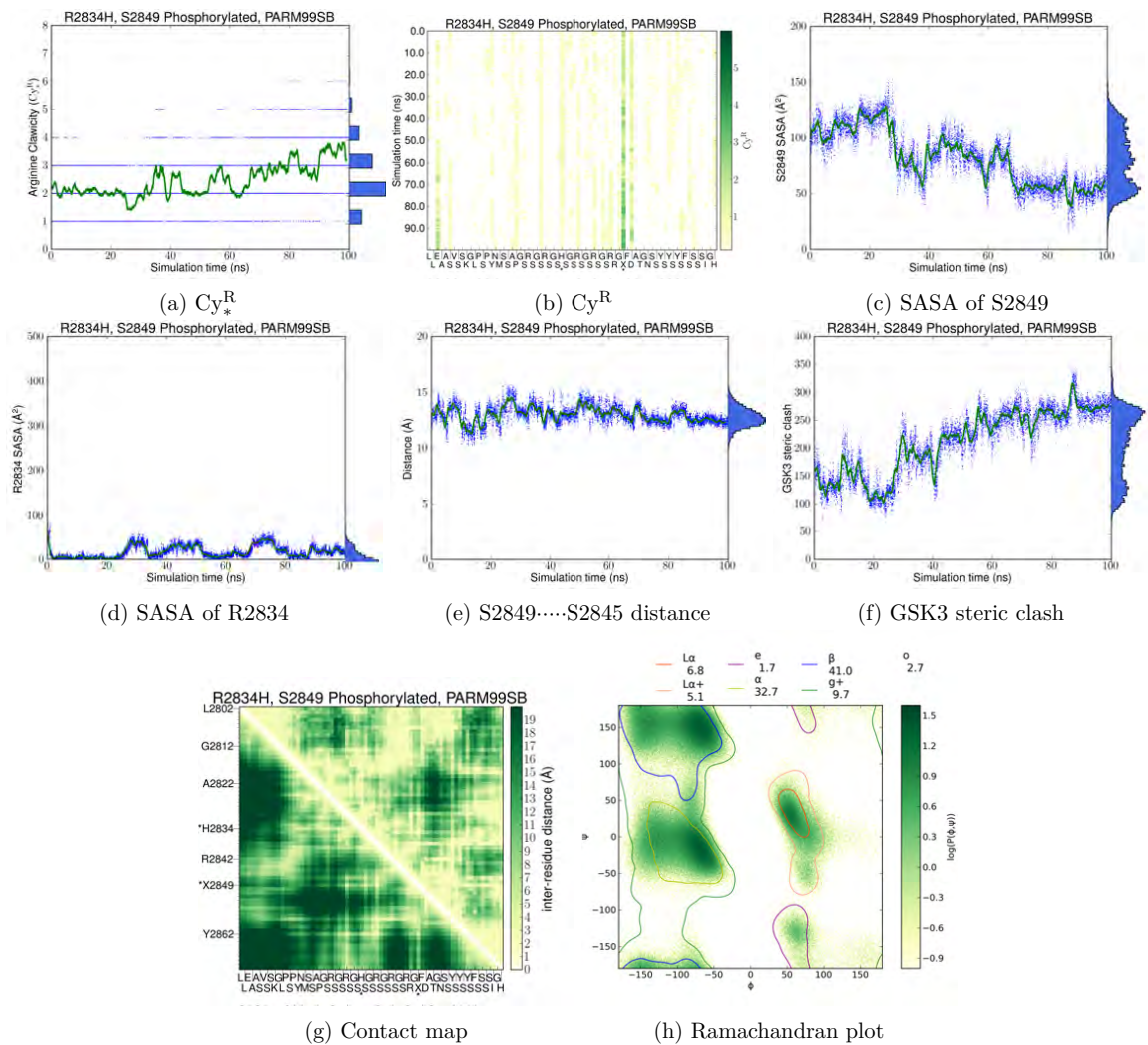

Figure S10: Behavior of **R2834H_S2849S2P_PARM99SB**: Time-series plots mark each observation as a blue point and contain a 1-ns running average as a green trace, and the marginal distributions are shown on the right axis, in each of panels a, c, d, e, and f.

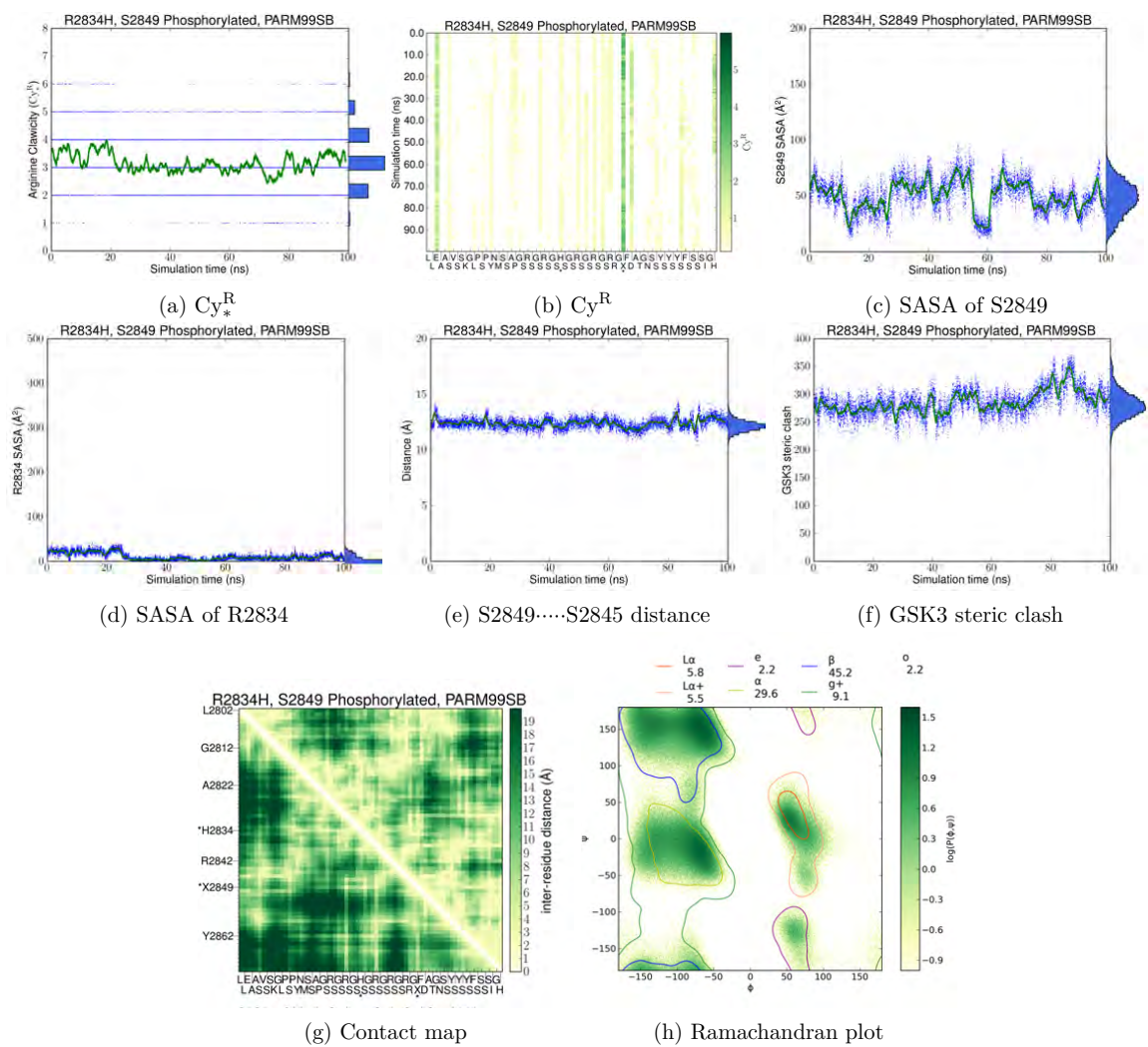

Figure S11: Behavior of **R2834H_S2849S2P_PARM99SB_cycle2**: Time-series plots mark each observation as a blue point and contain a 1-ns running average as a green trace, and the marginal distributions are shown on the right axis, in each of panels a, c, d, e, and f.

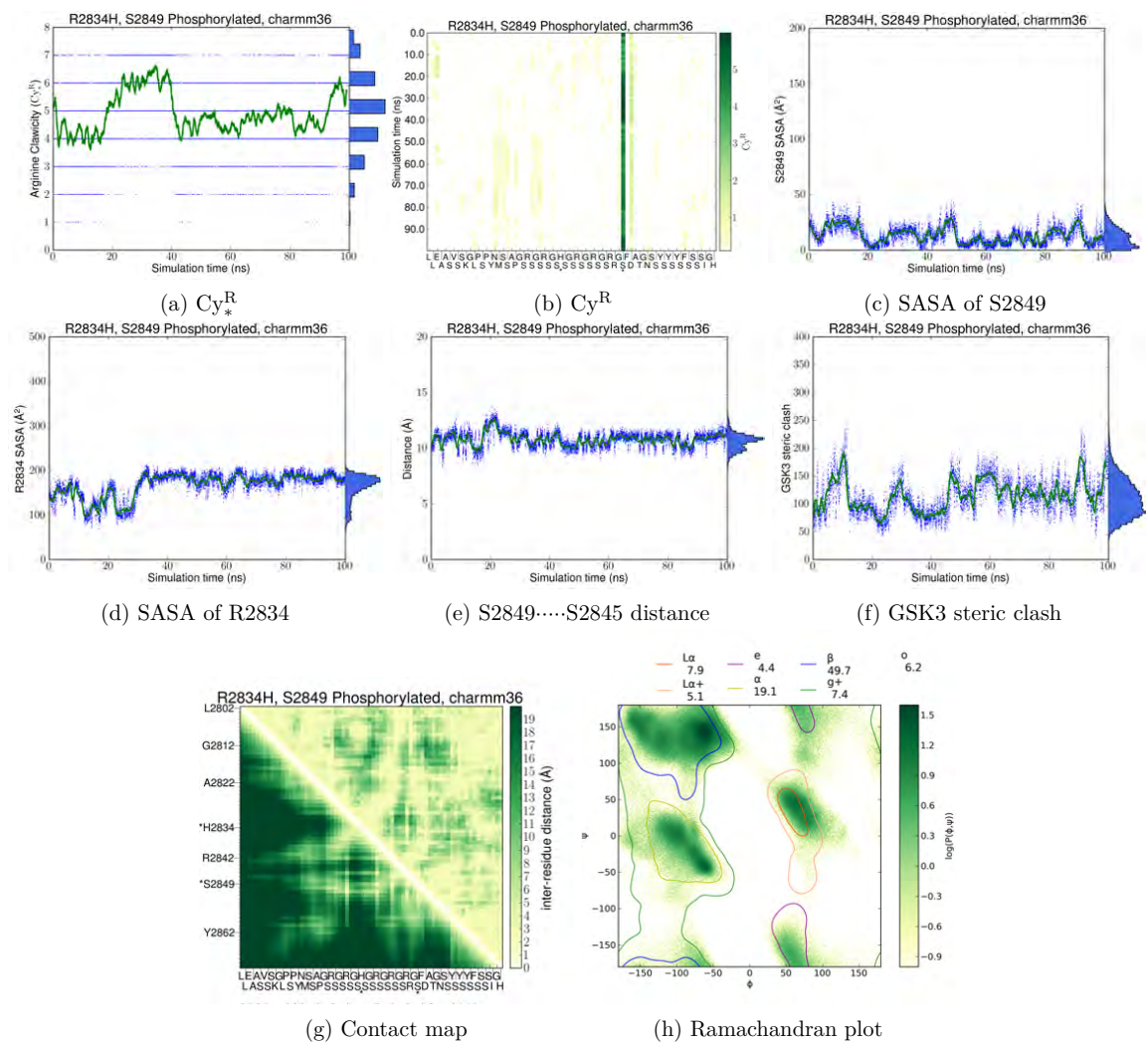

(a) $Cy^R_*$    (b) $Cy^R$    (c) SASA of S2849

(d) SASA of R2834    (e) S2849⋯S2845 distance    (f) GSK3 steric clash

(g) Contact map    (h) Ramachandran plot

Figure S12: Behavior of **R2834H_S2849S2P_CHARMM36**: Time-series plots mark each observation as a blue point and contain a 1-ns running average as a green trace, and the marginal distributions are shown on the right axis, in each of panels a, c, d, e, and f.

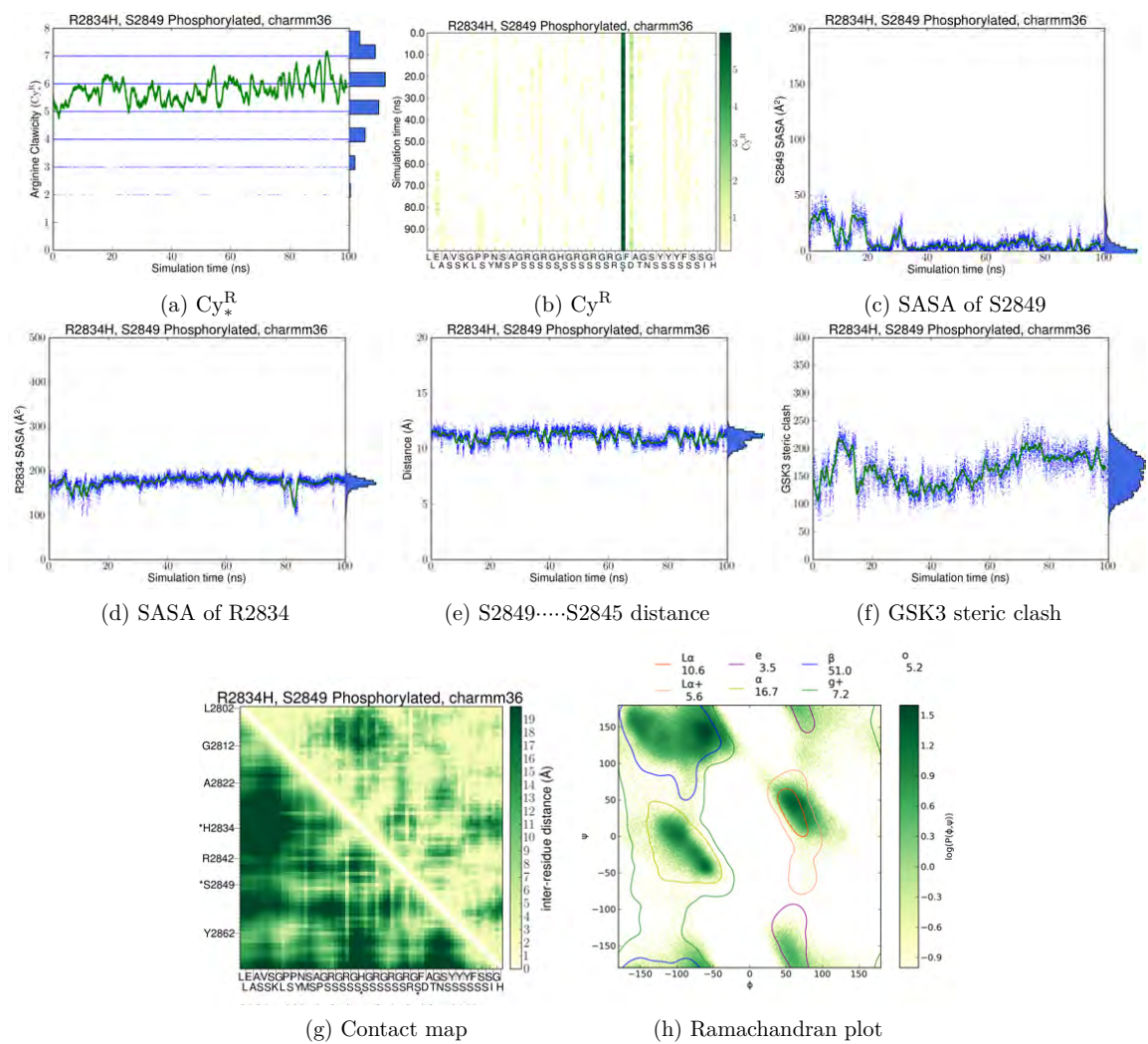

(a) $Cy^R_*$  (b) $Cy^R$  (c) SASA of S2849

(d) SASA of R2834  (e) S2849·····S2845 distance  (f) GSK3 steric clash

(g) Contact map  (h) Ramachandran plot

Figure S13: Behavior of **R2834H_S2849S2P_CHARMM36_cycle2**: Time-series plots mark each observation as a blue point and contain a 1-ns running average as a green trace, and the marginal distributions are shown on the right axis, in each of panels a, c, d, e, and f.

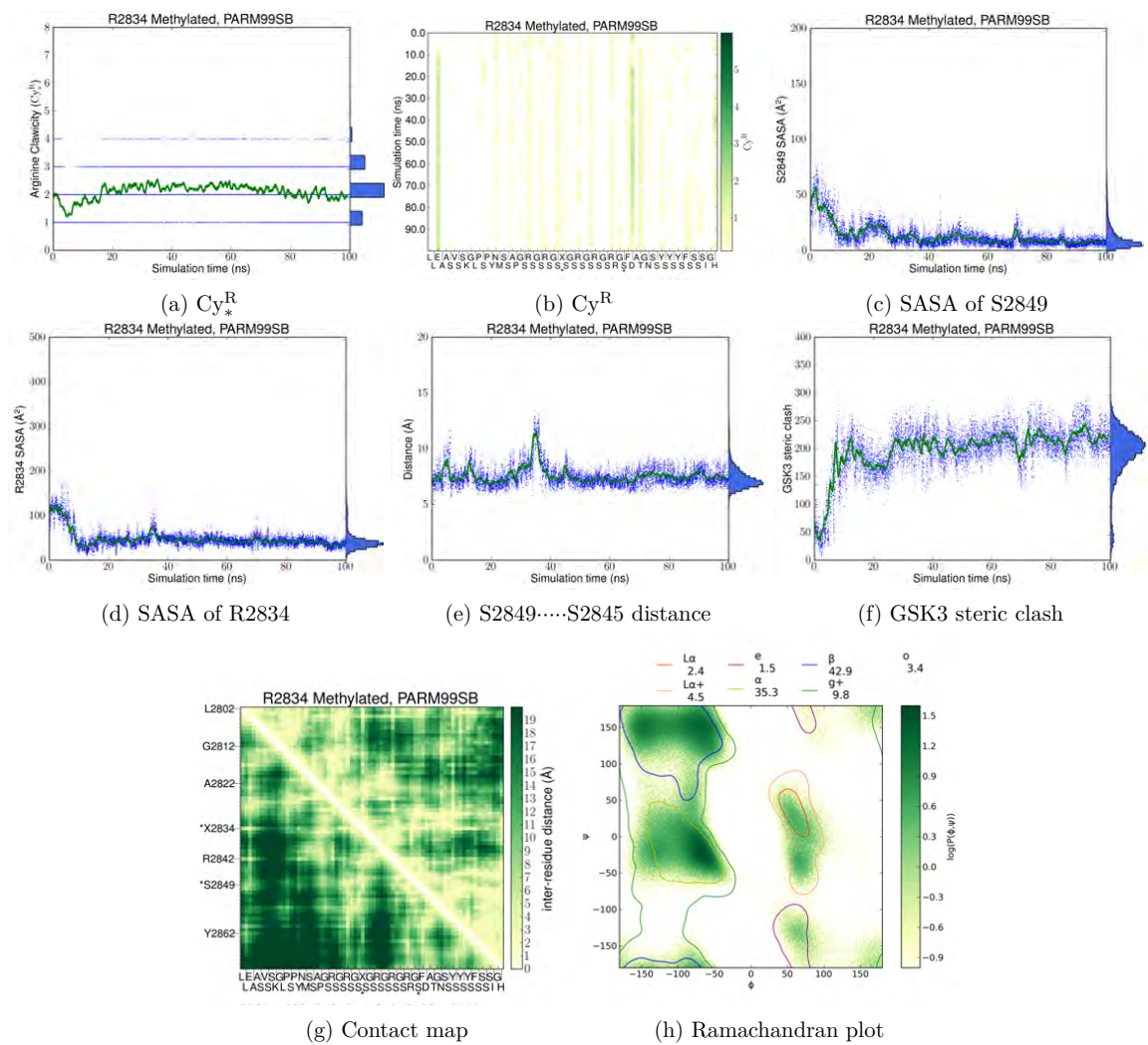

Figure S14: Behavior of **R2834MeMe_PARM99SB**: Time-series plots mark each observation as a blue point and contain a 1-ns running average as a green trace, and the marginal distributions are shown on the right axis, in each of panels a, c, d, e, and f.

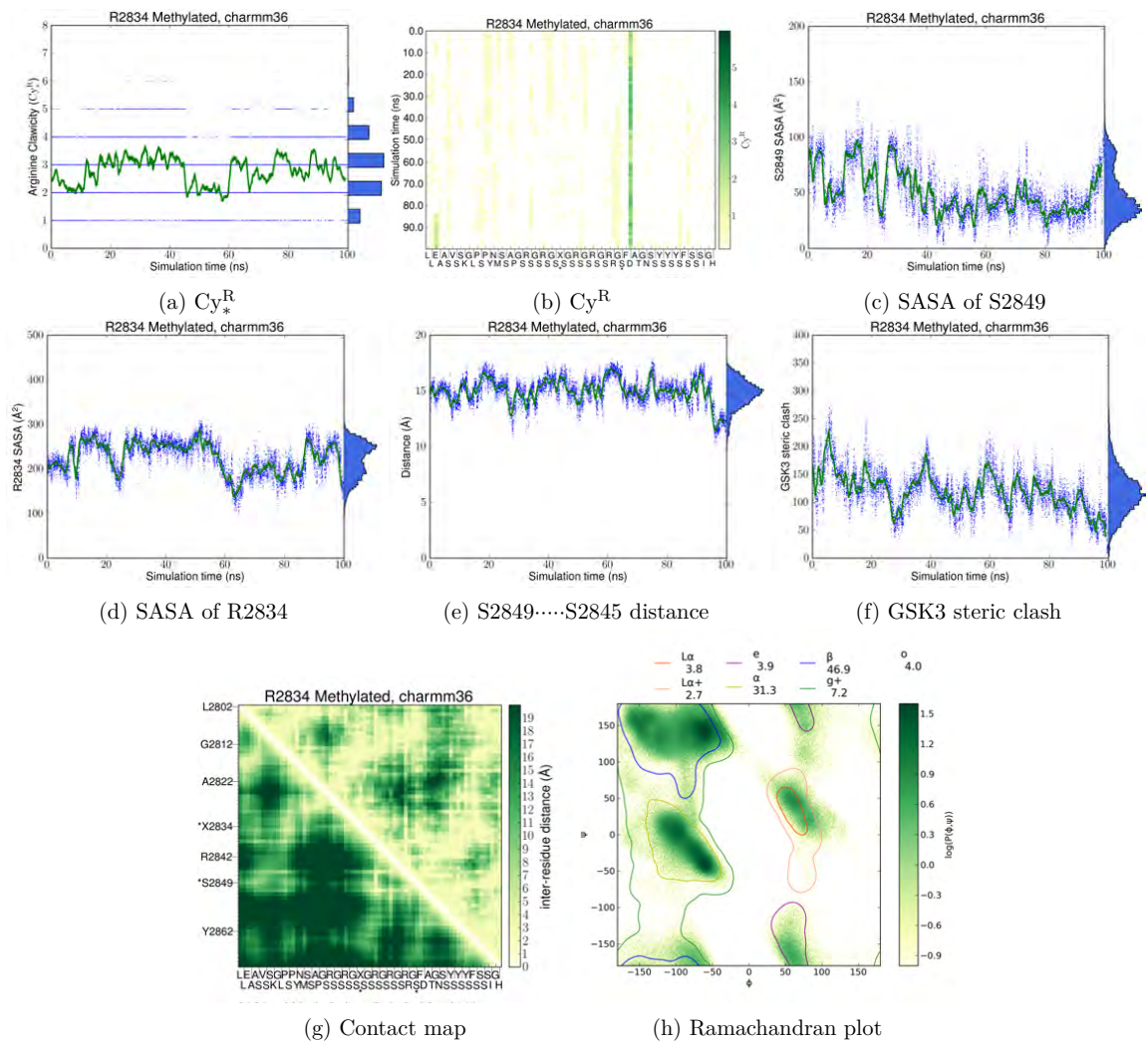

Figure S15: Behavior of **R2834MeMe_CHARMM36**: Time-series plots mark each observation as a blue point and contain a 1-ns running average as a green trace, and the marginal distributions are shown on the right axis, in each of panels a, c, d, e, and f.

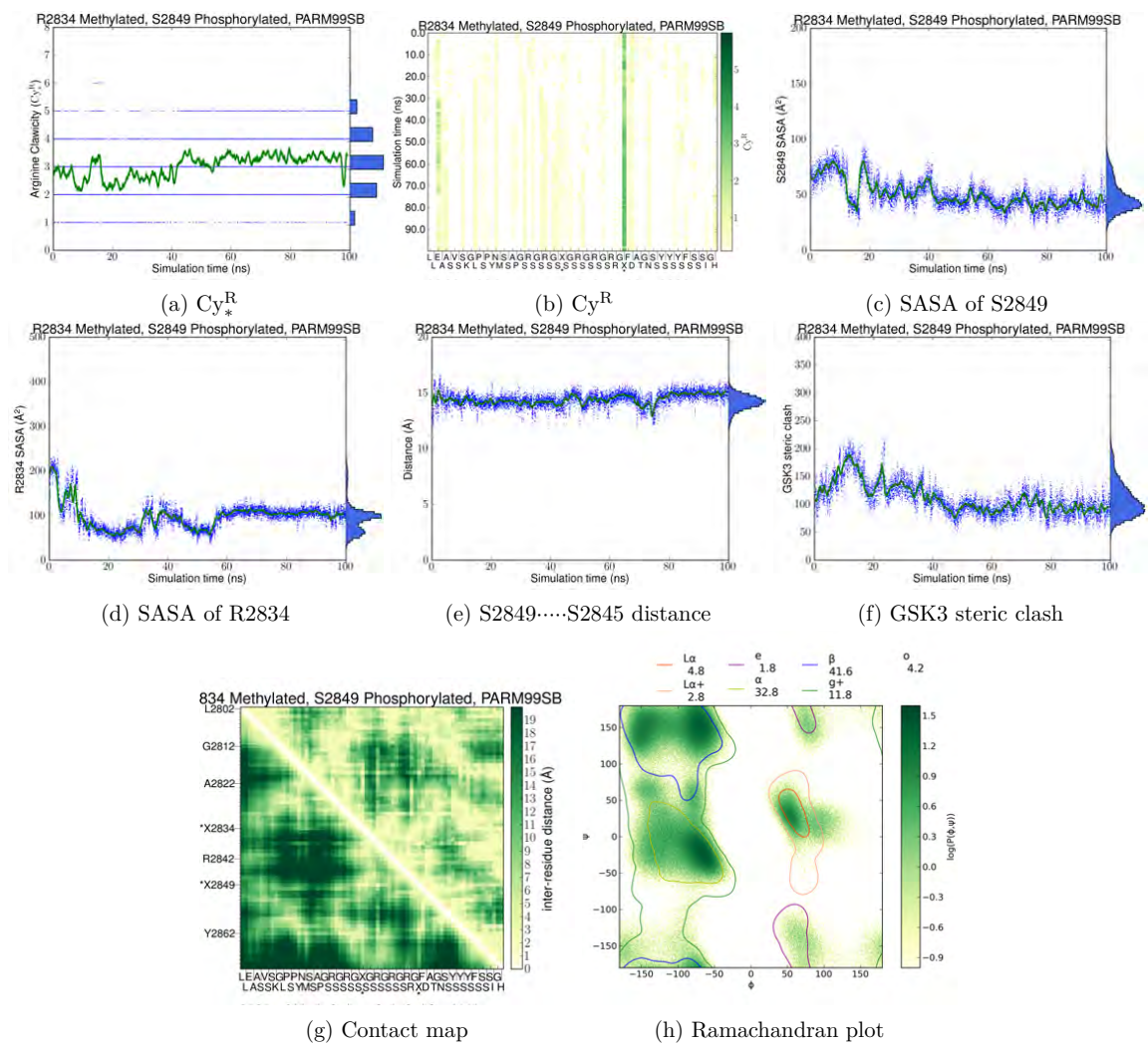

Figure S16: Behavior of **R2834MeMe_S2849S2P_PARM99SB**: Time-series plots mark each observation as a blue point and contain a 1-ns running average as a green trace, and the marginal distributions are shown on the right axis, in each of panels a, c, d, e, and f.

S18

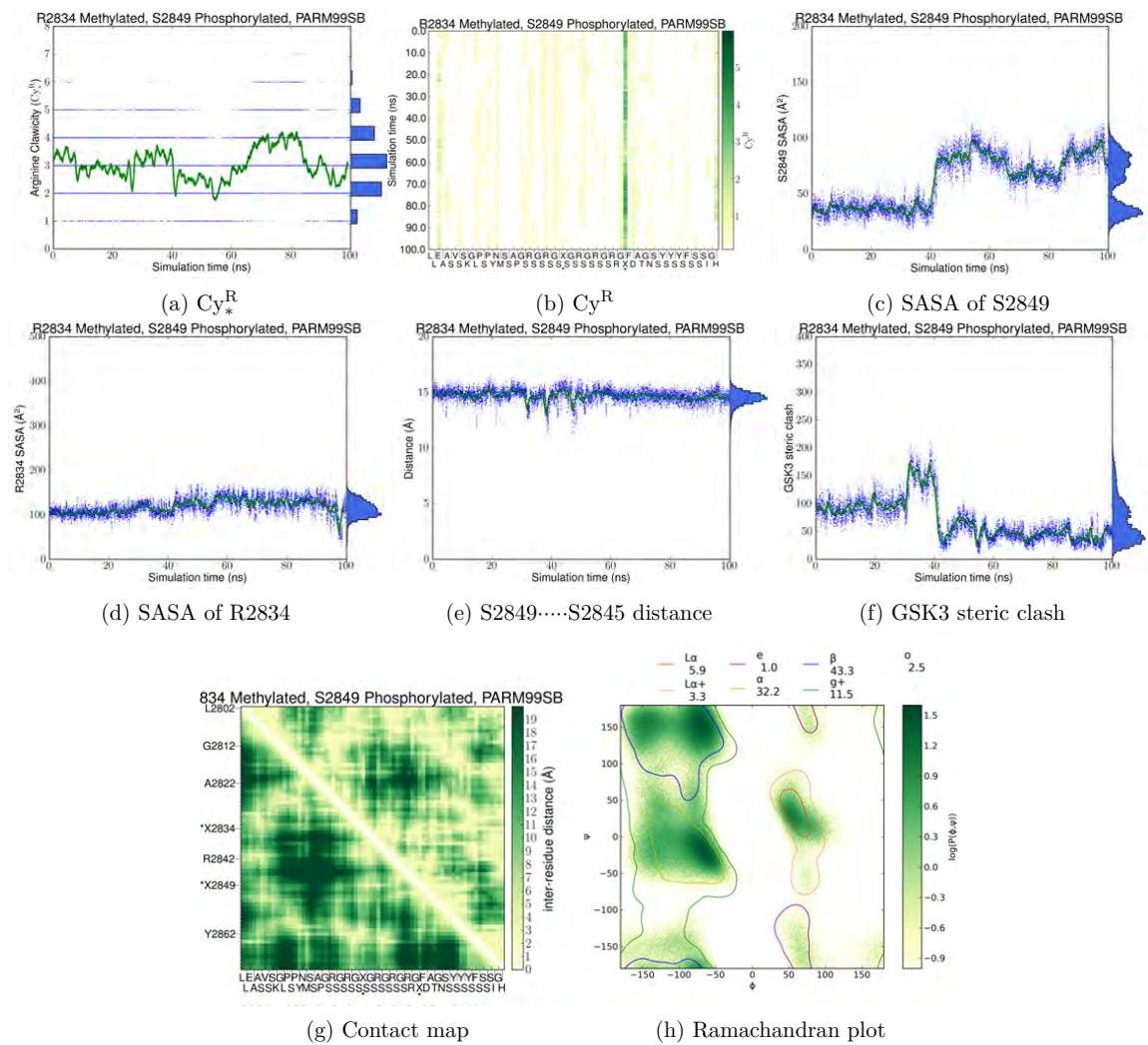

Figure S17: Behavior of **R2834MeMe_S2849S2P_PARM99SB_cycle2**: Time-series plots mark each observation as a blue point and contain a 1-ns running average as a green trace, and the marginal distributions are shown on the right axis, in each of panels a, c, d, e, and f.

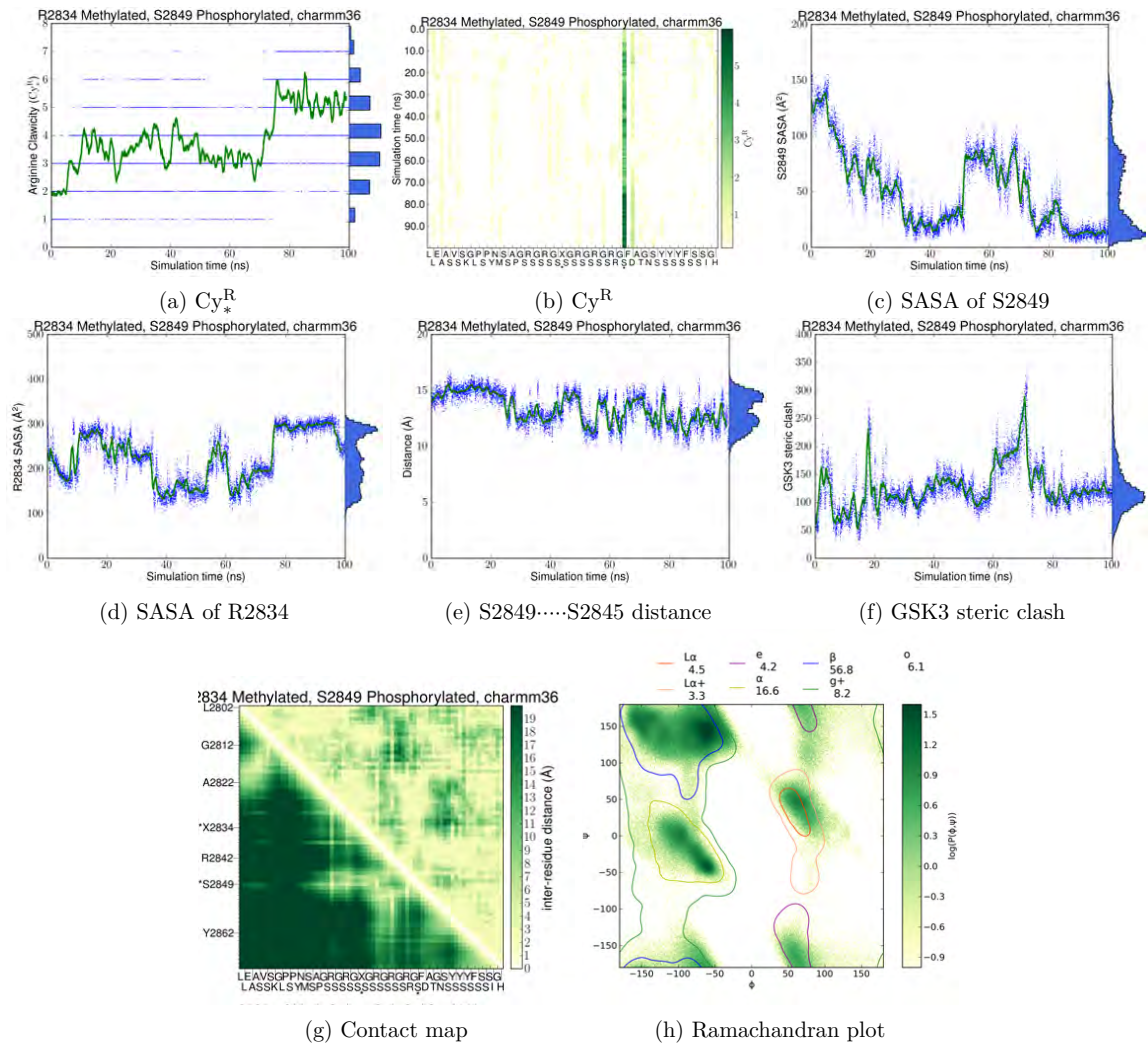

(a) $Cy^R_*$
(b) $Cy^R$
(c) SASA of S2849
(d) SASA of R2834
(e) S2849⋯S2845 distance
(f) GSK3 steric clash
(g) Contact map
(h) Ramachandran plot

Figure S18: Behavior of **R2834MeMe_S2849S2P_CHARMM36**: Time-series plots mark each observation as a blue point and contain a 1-ns running average as a green trace, and the marginal distributions are shown on the right axis, in each of panels a, c, d, e, and f.

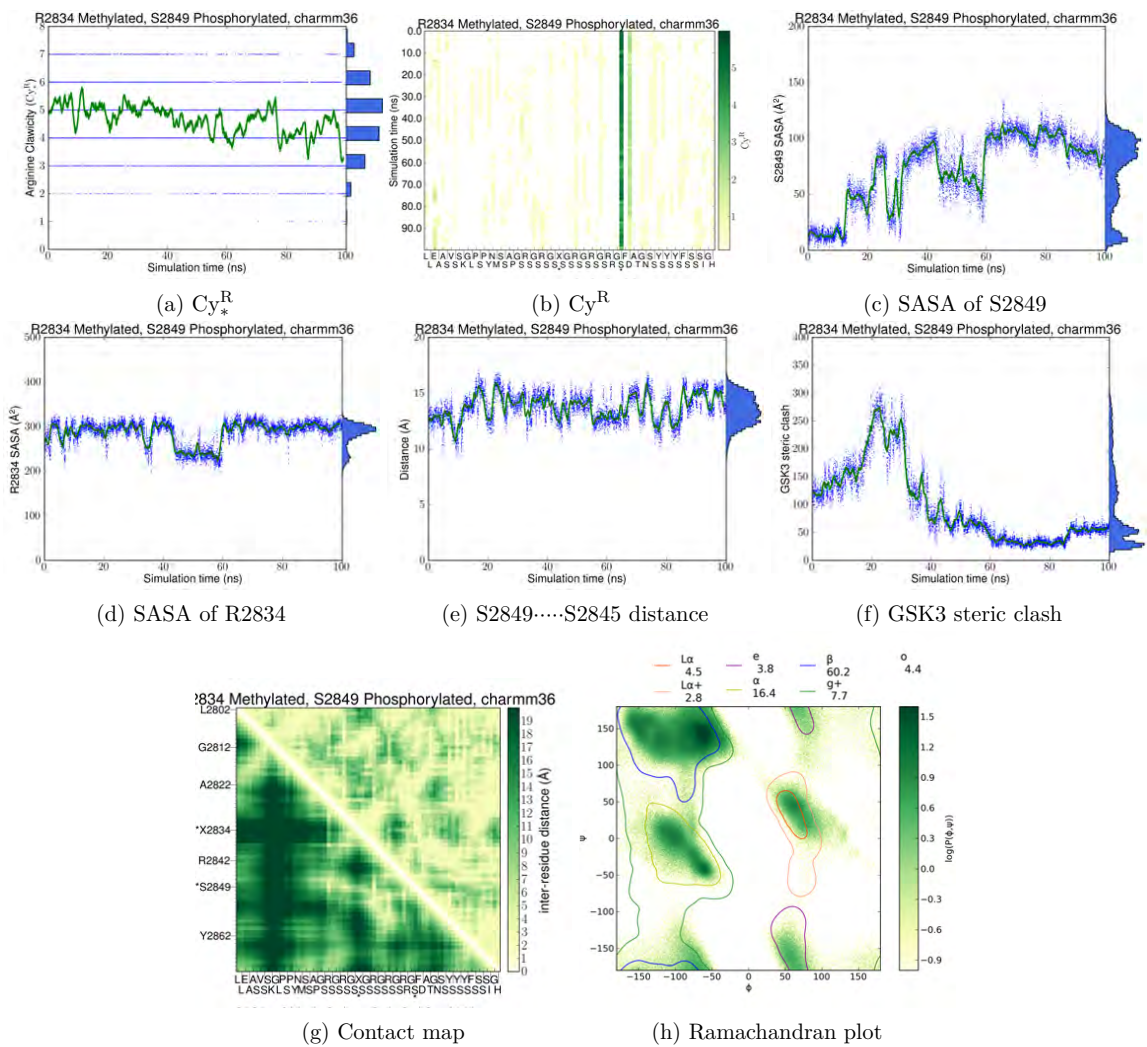

(a) $Cy^R_*$ (b) $Cy^R$ (c) SASA of S2849
(d) SASA of R2834 (e) S2849⋯S2845 distance (f) GSK3 steric clash
(g) Contact map (h) Ramachandran plot

Figure S19: Behavior of **R2834MeMe_S2849S2P_CHARMM36_cycle2**: Time-series plots mark each observation as a blue point and contain a 1-ns running average as a green trace, and the marginal distributions are shown on the right axis, in each of panels a, c, d, e, and f.